\documentclass[preprint,amsmath,amssymb]{revtex4-2}
\usepackage{graphicx}
\usepackage{dcolumn}
\usepackage{bm}
\usepackage{multirow}
\usepackage{float}
\usepackage{wrapfig}
\setcitestyle{super}
\usepackage[normalem]{ulem}
\usepackage{color}
\usepackage{xr}
\usepackage{array}
\usepackage{siunitx}
\usepackage{hyperref}
\usepackage{xr-hyper}
\usepackage{cleveref}
\hypersetup{
    colorlinks = true,
    urlcolor   = blue,
    citecolor  = blue,
    linkcolor = blue, 
}

\makeatletter
\providecommand \@ifxundefined [1]{%
 \@ifx{#1\undefined}
}%
\providecommand \@ifnum [1]{%
 \ifnum #1\expandafter \@firstoftwo
 \else \expandafter \@secondoftwo
 \fi
}%
\providecommand \@ifx [1]{%
 \ifx #1\expandafter \@firstoftwo
 \else \expandafter \@secondoftwo
 \fi
}%
\providecommand \href@noop [0]{\@secondoftwo}%
\providecommand \href [0]{\begingroup \@sanitize@url \@href}%
\providecommand \@href[1]{\@@startlink{#1}\@@href}%
\providecommand \@@href[1]{\endgroup#1\@@endlink}%
\providecommand \@sanitize@url [0]{\catcode `\\12\catcode `\$12\catcode
  `\&12\catcode `\#12\catcode `\^12\catcode `\_12\catcode `\%12\relax}%
\providecommand \@@startlink[1]{}%
\providecommand \@@endlink[0]{}%
\providecommand \url  [0]{\begingroup\@sanitize@url \@url }%
\providecommand \@url [1]{\endgroup\@href {#1}{\urlprefix }}%
\providecommand \urlprefix  [0]{URL }%
\providecommand \selectlanguage [0]{\@gobble}%
\providecommand \bibinfo  [0]{\@secondoftwo}%
\providecommand \bibfield  [0]{\@secondoftwo}%
\providecommand \BibitemShut  [1]{\csname bibitem#1\endcsname}%
\let\auto@bib@innerbib\@empty

\begin{document}

\preprint{}

\title{Conditions for the Emergence of Spontaneous Phonon Frequency Combs from Anharmonic Potentials}

\author{Jayakrishnan SS$^{1}$}

\author{Mayanak K Gupta$^{2,3}$}

\author{Dipanshu Bansal$^{1}$}

\affiliation{$^1$Indian Institute of Technology Bombay, Mumbai, MH 400076, India}
\affiliation{$^2$Homi Bhabha National Institute, Anushaktinagar, Mumbai MH 400094, India}
\affiliation{$^3$Solid State Physics Division, Bhabha Atomic Research Center, Mumbai, MH 400085, India}

\date{\today}
\begin{abstract}

Phononic frequency combs have been demonstrated experimentally and theoretically in the kHz-MHz regime under nonlinear driving; however, their spontaneous formation in the GHz-THz regime remains rare. We investigate the spontaneous formation of frequency combs in van der Waals solid CrGeTe$_3$, where combs of spacing $\sim$2\,cm$^{-1}$ were experimentally proposed to occur in a flat phonon mode but are now reported to originate from isotopic distribution. Our spectral energy density calculations using machine-learning-augmented molecular dynamics simulations show comb-like features with reduced spacings of $\sim$0.1 and 0.6\,cm$^{-1}$ in the same mode at 50 and 200\,K. However, the spacings of the comb-like features are comparable to the phonon linewidth, making precise identification of the frequency comb challenging. From a comprehensive analysis of different modes of CrGeTe$_3$, we identified two conditions to distinctively observe the combs beyond the experimental resolution: (i) an intramolecular or isolated mode with reasonably large third- or higher-order anharmonicity from phonon self-interaction and (ii) limited phonon scattering channels. Our analysis of other van der Waals solids highlighted that these conditions are met for the nearly flat intramolecular $E_{2g}$ and $A_{2u}$ phonon modes of WSe$_2$. The detailed calculations showed the formation of a spontaneous frequency comb with spacings of 1.7 and 0.67\,cm$^{-1}$ from coherent superposition of four and five-phonon eigenstates in the $E_{2g}$ and $A_{2u}$ phonon modes, respectively. Available Raman scattering data of the $E_{2g}$ mode in the literature provide preliminary, if not conclusive, evidence of comb formation. Our findings offer a deeper understanding and open avenues for engineering long-lived phononic frequency combs for various practical applications. 

\end{abstract}

\maketitle
\section{Introduction}
Optical frequency combs, consisting of a series of evenly spaced lines, are routinely used in precision frequency measurements, accurate timekeeping, molecular sensing, distance measurements, attosecond pulse generation, and spectroscopic analysis.~\cite{pupeza2021extreme,kippenberg2011microresonator,udem2002optical,cundiff2003colloquium,gohle2005frequency,jones2005phase,keilmann2004time,huang2008spectral} While frequency combs are typically associated with nonlinear optical processes, theoretical studies have shown the emergence of phonon frequency combs in the Fermi-Pasta-Ulam-Tsingou $\alpha$ chains.~\cite{cao2014phononic} Recently, experimental demonstration of phononic frequency combs using these $\alpha$ chains in the kHz and MHz regime was realized by piezoelectrically driven micro-mechanical resonators and carbon nanotube mechanical resonators.~\cite{ganesan2017phononic,ganesan2018phononic,ganesan2018phononicprb,xu2025phononic} Such phononic combs have applications in phonon computing, accurate micro and nanomechanical resonant sensors, and phase coherent phonon lasers.~\cite{wang2007thermal,middlemiss2016measurement,vahala2009phonon} Despite advances in theoretical and experimental optical techniques, evidence of phononic frequency combs in the GHz-THz regime is limited. Recent theoretical studies showed the generation of THz-frequency combs from nonlinear coupling between Higgs-like and Goldstone-like phonon modes in hexagonal InMnO$_3$.~\cite{rangwala2026spontaneous} Experimentally, Raman measurement on CrGeTe$_3$ exhibited spontaneous formation (not in the resonators) of discrete energy spacing of approximately 2\,cm$^{-1}$ for the $A_g(5)$ mode, which the authors theorized to originate from cubic anharmonic lattice potential.~\cite{chen2025spontaneously} However, these spontaneous frequency combs are now understood to originate from the natural isotopic mass distribution of Ge instead of combing.~\cite{krasucki2025spin,mayur1994fine} A similar observation is now reported on InSiTe$_3$,~\cite{belojica2026phonon} which can likewise be explained by the natural isotopic mass distribution of Si rather than true phonon comb formation. Hence, understanding how spontaneous phonon frequency combs in GHz and THz regimes arise and under what conditions they remain observable remains an open question.

The idea of generating spontaneous phonon frequency combs from higher-order anharmonicity by Chen \emph{et al.}~\cite{chen2025spontaneously} is promising, and warrants further exploration of the conditions under which they are observable. Anharmonicity manifests as higher-order terms in the phonon potential. In the absence of cubic or higher-order terms, the potential reduces to a harmonic phonon potential, characterized by a constant spacing between different phonon eigenstates $E_n$ and $E_{n+1}$ (see Fig.~\ref{fig_figure1}d), and a single spectral line in spectroscopy measurements. In contrast, a cubic, quartic, or higher-order potential will induce varying spacings between the phonon eigenstates $E_n$ and $E_{n+1}$, and coherent superposition of these states 
can lead to parametric oscillation and formation of phononic frequency combs,~\cite{chen2025spontaneously} as shown in Fig.~\ref{fig_figure1}e. A phenomenological description of these parametric oscillations and phononic frequency combs is provided by Chen \emph{et al.}~\cite{chen2025spontaneously} Note that the spacing between frequency combs remains constant for cubic phonon potential (i.e., $A_n-A_{n-1} = A_{n+1}-A_{n}$, see Fig.~\ref{fig_figure1}e for notation), and increases or decreases for quartic or higher-order potentials depending on the sign of higher-order coefficients. Even in high-spectral-resolution experiments, such as Raman scattering, these frequency combs may appear as a single spectral feature if various phonon scattering channels are significant enough to broaden the spectral feature to more than the spacing between them. For example, in Fig.~\ref{fig_figure1}f, if the phonon linewidth $\Gamma$ from scattering is more than the spacing $A_n-A_{n-1}$, then instead of frequency combs, a single broadened feature will be visible. It is worth noting that large losses leading to decoherence make the identification of such features as frequency combs increasingly challenging, and in the limit of complete decoherence, the frequency combs loses its meaning entirely. However, if the phonon scattering channels are few, the coherent superposition of discrete states may lead to parametric oscillation, forming spontaneous phonon frequency combs.~\cite{chen2025spontaneously} This delicate balance of anharmonicity will generally be observable in intramolecular or isolated vibrations with anharmonicity governed by phonon self-interaction, rather than phonon-phonon interaction, as the latter will also lead to finite $\Gamma$ (see Supplementary Information (SI) Section~\ref{anharmonicity} for more details). Moreover, the spontaneously formed combs will have a higher probability of forming at low temperatures, as the phonon-phonon scattering will be relatively low,~\cite{Born1988, Bruesch, Dove1993} and can be observed using high-resolution spectroscopy techniques. However, spontaneous frequency combs in optical phonons have not yet been reported in the literature, as optical phonons tend to decay in a few cycles due to strong mode-mode coupling.~\cite{Born1988, Bruesch, Dove1993, wu2008ultrafast} 

Two-dimensional  (2D) van der Waals materials have the potential to serve as an excellent platform for exploring frequency comb phenomena. For example, similar to piezoelectrically driven micro-mechanical resonators and carbon nanotube mechanical resonators,~\cite{ganesan2017phononic,ganesan2018phononic,ganesan2018phononicprb,xu2025phononic} recently, phonon frequency combs in nanoscale resonators have been theoretically demonstrated in twisted MoS$_2$ using fixed-frequency and rapid frequency-sweep through molecular dynamics simulations, highlighting the capability of the simulation technique to probe and analyze frequency combs.~\cite{liu2025phononic} We must emphasize that while the optical cavity and micro/nano-resonator techniques allow fewer decay paths to optical pulses and acoustic phonons, respectively, and consequently the observation of frequency combs; however, the spontaneous formation of frequency combs from THz optical phonons is rare.~\cite{chen2025spontaneously} Our study is exclusively focused on whether CrGeTe$_3$ could be a potential candidate to observe spontaneous frequency combs. If not, what conditions are necessary for observation of these combs, and is there another material where combs may be observed and detected? In this study, we provide a quantitative description of spontaneous frequency comb formation using density-functional-theory-based machine-learning-augmented molecular dynamics (MLMD) simulations and phenomenological modeling in two prototypical 2D materials, CrGeTe$_3$ and WSe$_2$. 

\section{Results and Discussions}

\begin{figure}
    \centering
    \includegraphics[trim=2cm 0.5cm 1.2cm 2.8cm, clip=true,width=0.75\linewidth]{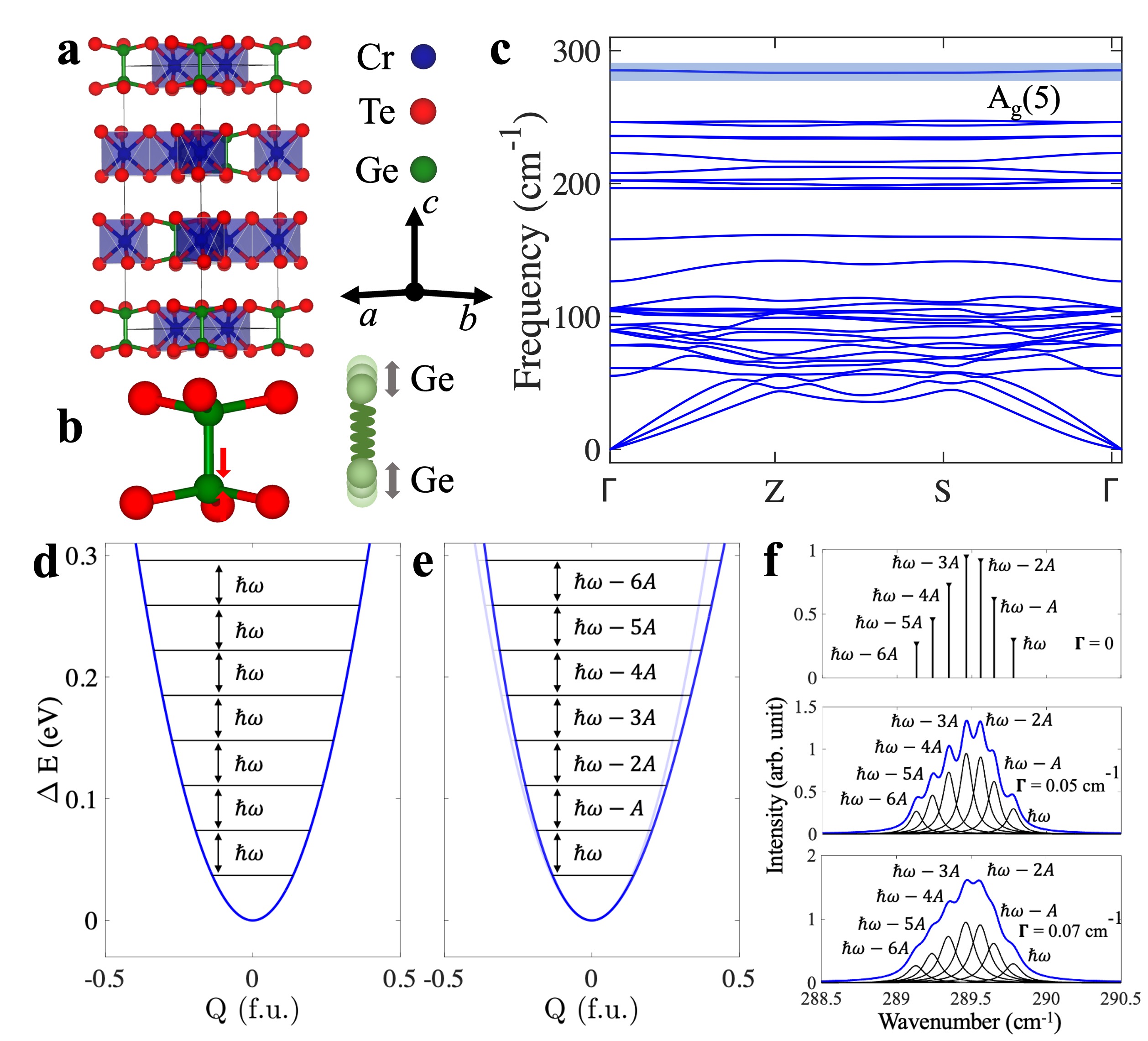}
    \caption[Structure of CrGeTe$_3$]{(a) Crystal structure of CrGeTe$_3$ showing CrTe$_6$ octahedra in light blue. (b) The phonon eigenvector of the $A_g(5)$ mode showing internal compression/stretching motion of the Ge-Ge pair. Schematic representation of the Ge-Ge dimer, approximated as an isolated oscillator, is also shown. (c) Phonon dispersion of CrGeTe$_3$. The spatially localized isolated flat $A_g(5)$ phonon mode is highlighted. (d,e) Comparison of harmonic (quadratic) and anharmonic (higher-order) potentials and their corresponding vibrational eigenstates. Harmonic potential (light blue) is shown alongside the anharmonic potential (dark blue) for comparison. Equally spaced eigenstates in the harmonic oscillator lead to a single transition energy at $\hbar\omega$, while cubic anharmonicity results in unequally spaced eigenstates with spacing reduction by $A$ in each higher energy eigenstate. (f) Difference between eigen states ($\hbar\omega$) of the cubic potential with no phonon broadening (i.e., $\Gamma = 0$, top panel), phonon broadening with $\Gamma$ of 0.05 and 0.07\,cm$^{-1}$ (middle and bottom panels). Large broadening leads to masking of discrete frequency combs.}
 \label{fig_figure1}
\end{figure}

CrGeTe$_3$ crystallizes in a rhombohedral $R\overline{3}$ space group (No. 148) (see Fig.~\ref{fig_figure1}a) and undergoes a transition to a ferromagnetic state below the Curie temperature of  $\sim$65\,K.~\cite{carteaux1995crystallographic} It harbors five $E_g$ and five $A_g$ Raman-active modes.~\cite{chen2025spontaneously} Raman spectroscopy for the $A_g(5)$ mode showed multiple phonon peaks (similar to a frequency comb), which is now understood to originate from the isotope mass effect.~\cite{chen2025spontaneously,krasucki2025spin} This $A_g(5)$ mode involves the vibration of Ge atoms moving toward each other (see Fig.~\ref{fig_figure1}b) with a flat dispersion in the entire Brillouin zone (see Fig.~\ref{fig_figure1}c), indicating that the mode is spatially localized and weakly coupled to other phonons.~\cite{chen2025spontaneously,krasucki2025spin,zhang2019first} 

First, we focus on whether the $A_g(5)$ mode is a potential candidate for harboring frequency combs. As mentioned previously, for combs to be visible, the isolated mode must have relatively large magnitudes of cubic and/or quartic terms and minimum scattering channels to distinctly observe the combs. 
When solving the Schrödinger equation for a phonon of energy $\sim$36\,meV (same energy as $A_g(5)$ mode) having a cubic phonon potential (normalized to maximum amplitude $Q_D = 0.4$\,\AA), we find that the ratio of the cubic to the quadratic terms ($k_3/k_2$) needs to be above 27\% for comb spacing to be larger than the typical accessible spectral resolution of 0.4\,cm$^{-1}$, i.e., $A_n-A_{n-1}>0.4$\,cm$^{-1}$. Similarly, for a symmetric potential for a phonon of the same energy, where cubic terms drop out, the quartic to the quadratic term ratio ($k_4/k_2$) needs to be above 9.5\%. For modes with energies above and below $\sim$36\,meV, $k_3/k_2$ and $k_4/k_2$ will be smaller and larger, respectively. To put the anharmonicity magnitude in perspective, we compare these terms' ratio to terms obtained for the unstable $H_1$ and $K_5$ modes driving the concomitant structural and magnetic phase transition in h-FeS; for the symmetric $H_1$ mode with $Q_D^{H_1} = 1.14$\,\AA, $k_4/k_2$ is  $\sim$50\%, while for the $K_5$ mode of $Q_D^{K_5} = 0.35$\,\AA, $k_3/k_2$ and $k_4/k_2$ is $\sim$15\%.~\cite{bansal2020magnetically}  Hence, the need for large $k_3/k_2$ and $k_4/k_2$ ratios for the combs to be experimentally observable implies the phonon potential must be strongly anharmonic unless the phonon energies are quite high (of the order of 100\,meV), where small ratios can provide comb spacing of 0.4\,cm$^{-1}$. This anharmonicity should ideally arise from self-interaction rather than phonon-phonon interactions, which is generally true for intramolecular or isolated flat modes. Note that these modes differ from ideal Einstein modes (harmonic oscillators) as they must possess anharmonicity from self-interaction, which is precisely the requirement for spontaneous phonon frequency combs. We provide the distinction between anharmonicity from phonon self-interaction and phonon-phonon interaction in SI Section~\ref{anharmonicity}. 

\begin{figure}
    \centering
    \includegraphics[trim=0cm 0cm 0cm 0cm, clip=true,width=0.75\linewidth]{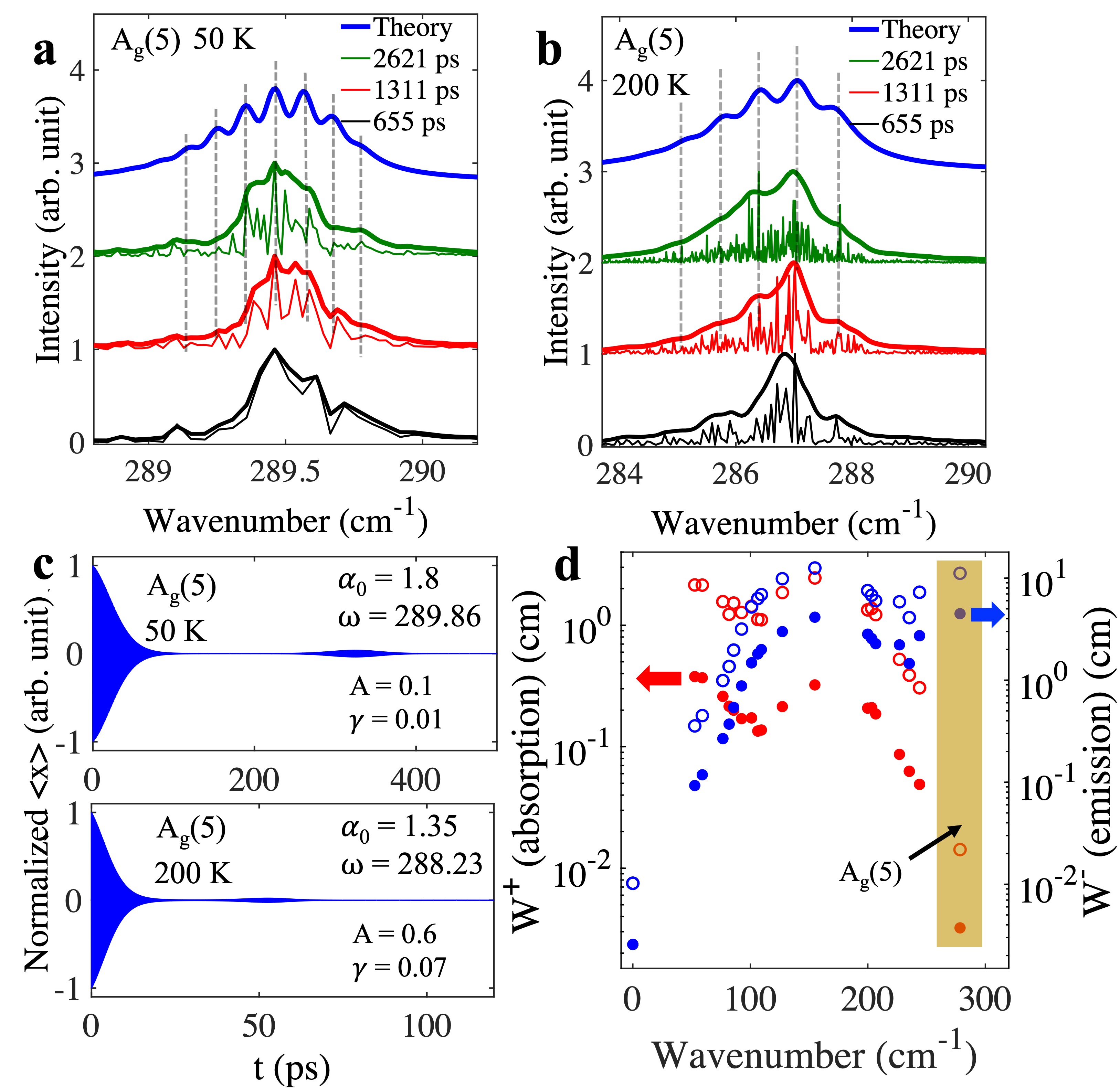}
    \caption[SED of CrGeTe$_3$]{(a) Comb-like features of the $A_g(5)$ mode of CrGeTe$_3$ as obtained by calculating SED from MLMD simulation at 50\,K for different trajectory lengths and phenomenological model (referred to as theory). The spectra are vertically offset. The thick lines above the raw SED data show the spectrum convolved with a Lorentzian with FWHM $=0.025$\,cm$^{-1}$ for ease of visualization (comparison with other FWHM values is shown in SI Fig.~\ref{suppfig29_differentbroadening_200K}). Black dashed lines indicate the peak position of combs. (b) Same as panel (a) but for 200\,K convoluted with a Lorentzian having FWHM $=0.15$\,cm$^{-1}$. (c) $\langle x \rangle_t$ obtained from the phenomenological model. (d) Three phonon scattering phase space -- absorption ($W^{(+)}$) and emission ($W^{(-)}$) -- of phonon modes at the zone-center at 50\,K (filled markers) and 200\,K (empty markers).}
        \label{fig_CGTphononspectra}
\end{figure}

Based on the above discussion, we now quantitatively estimate the phonon-phonon and spin-phonon scattering of the flat dispersion $A_g(5)$ mode. Electron-phonon scattering is expected to be negligible since the band gap is $\sim$0.37\,eV (see SI Fig.~\ref{fig_bandstructure}), and is not discussed further. Here, we must emphasize that spatially localized vibrations leading to a flat dispersion mode in the Brillouin zone do not necessarily imply the absence of phonon scattering, thus necessitating the need for explicit calculations of scattering phase space. Three-phonon scattering phase space can be calculated as, $W^{(\pm)}(\textbf{q}j) = \frac{1}{N_\textbf{q}} \sum_{\textbf{q}_1,\textbf{q}_2,j_1,j_2} {\begin{Bmatrix}n_2-n_1 \\ n_1+n_2+1 \end{Bmatrix}}\delta(\omega_{\textbf{q}j} \pm \omega_{\textbf{q}_1j_1}-\omega_{\textbf{q}_2j_2}) \delta_{\textbf{q} \pm \textbf{q}_1,\textbf{q}_2+\textbf{G}}$, where, \textbf{q} is the phonon wavevector, $n$ is the occupation factor, $\omega$ is the phonon frequency, and $j$ is the mode number (see SI Section~\ref{Scattering_phase_space}).~\cite{tadano2014anharmonic} Figure~\ref{fig_CGTphononspectra}d shows the absorption ($W^+$) and emission ($W^-$) channels of all modes at the zone-center at 50 and 200\,K. The response of the $A_g(5)$ mode is shaded for easy viewing. Here, we find that while $W^+$ for the $A_g(5)$ mode is negligible, $W^-$ is finite even at 50\,K, which further increases with temperature. The presence of finite magnitude in $W^-$ implies that the $A_g(5)$ mode will have a finite $\Gamma$. 
Similarly, we quantitatively evaluate the spin-phonon coupling using the Wannier orbitals-based Green's function method (see SI Sections \ref{spinphononcoupling}).~\cite{korotin2015calculation,jayakrishnan2025coherent,ss2025effect} SI Figure~\ref{fig_phononmagnonscattering}h shows that the three nearest neighbor exchange interactions do not exhibit any change with $A_g(5)$ mode perturbation, showing minimal spin-phonon coupling for the $A_g(5)$ mode, consistent with previous experimental and computational studies.~\cite{tian2016magneto,chen2025spontaneously,zhang2019first,krasucki2025spin}

\begin{figure}
    \centering
    \includegraphics[width=1\linewidth]{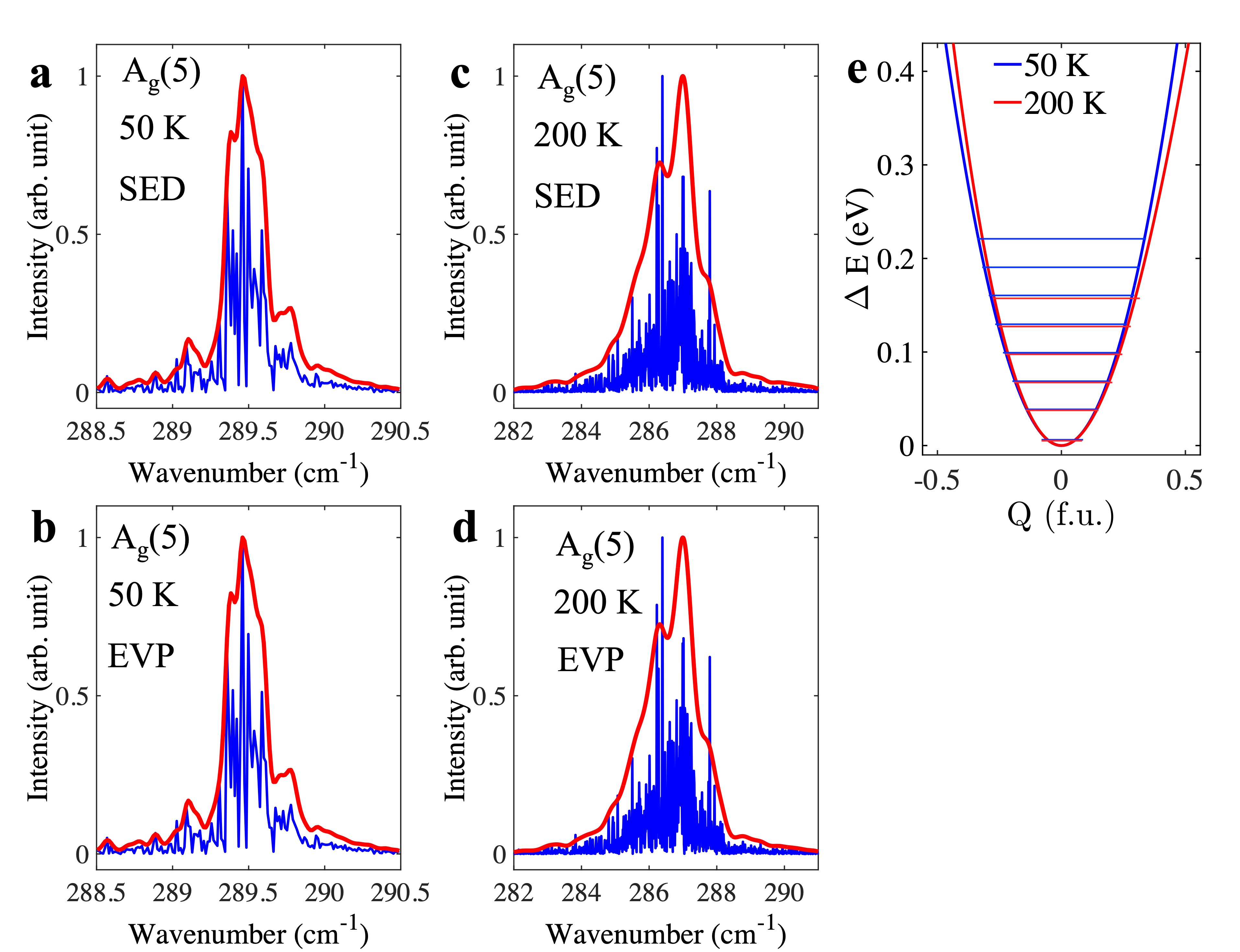}
    \caption[CGT SED and EVP Comparison]{(a-d) Comparison of spectral energy density (SED) and zero-Kelvin eigenvector-projected (EVP) spectrum for the $A_{g}(5)$ mode at 50 and 200\,K for trajectory length of 2621\,ps. The red curve in panels (a,b) is convolved with a Lorentzian having a FWHM of 0.025\,cm$^{-1}$, and panels (c,d) with a FWHM of 0.15\,cm$^{-1}$ for ease of visualization. (e) Comparison of potential of the $A_g(5)$ mode at 50\,K (blue curve) and 200\,K (red curve). The phonon eigenstates are shown within the potential. At higher energies, the difference between eigenstates of 50 and 200\,K increases due to the difference in the comb-like structure spacings. Here, 1\,f.u. corresponds to 0.40\,\AA.}
    \label{fig_CGTSEDEVPcomparison}
\end{figure}

From the above quantitative calculations, it is apparent that for the $A_g(5)$ mode, phonon-phonon scattering has a finite magnitude, but spin-phonon coupling is negligible. We now use our MLMD simulations (see SI Section~\ref{MLMDsimulationdetails}), which are uniquely suited to capture transitions between all energy eigenstates and their coherent superposition, to probe if spontaneous frequency combs exist for the $A_g(5)$ mode. Figure~\ref{fig_CGTphononspectra}(a,b) shows the spectral energy density (SED) at 50 and 200\,K in the vicinity of $A_g(5)$ mode for different trajectory lengths, providing spectral resolution of 0.051 to 0.013\,cm$^{-1}$. We convolute the raw SED data (thin solid line) with a Lorentzian of full-width-half-maximum (FWHM) of 0.025 and 0.15\,cm$^{-1}$ (thick solid line) for ease of visualization and comparison, showing several satellite peaks separated by 0.1 and 0.6 \,cm$^{-1}$, which remains independent of trajectory length. We attribute the increase in the spacing between peaks from 0.1 to 0.6\,cm$^{-1}$ with temperature to the change in phonon potential, which is largely harmonic at 50\,K ($ k_3/k_2 = 13.6\%$), but becomes increasingly anharmonic ($ k_3/k_2 = 33\%$) at 200\,K (see SI Fig.~\ref{fig_CGTSEDEVPcomparison}e). These peaks nearly merge into each other because of the finite $\Gamma$ expected from phonon-phonon scattering as discussed above. Note that for the trajectory length of 655\,ps, since spectral resolution is 0.051\,cm$^{-1}$, all peaks are not clearly resolved. To further ensure that all observed features originate from the same $A_g(5)$ mode, we projected the 0\,K eigenvector on the MLMD trajectory (see SI Fig.~\ref{fig_CGTSEDEVPcomparison}), which is nearly identical to SED calculations. Observation of satellite peaks originating from the same phonon mode suggests the formation of frequency comb-like features, but the finite phonon-phonon scattering nearly merges the features (these features are more conclusive for WSe$_2$ as discussed later).  

To understand the occurrence of comb-like features, we use the phenomenological model proposed by Chen \emph{et al.}~\cite{chen2025spontaneously} The expectation value of $x$ describing the classical motion of atoms within the phonon mode can be written as (see SI Section~\ref{Oscillatormodel}): 
\begin{equation} \label{expectationvalue_main}
\langle x \rangle_t = \alpha_0 \mathrm{exp}[-\iota (\omega-A)t + |\alpha_0|^2(e^{\iota A t}-1)]+c.c.
\end{equation}
Here, $A$ is positive and equal to the spacing between the adjacent peaks, $\omega$ is the frequency of the mode, $\alpha_0$ is a complex eigenvalue for the coherent state, and $|\alpha_0|$ is analogous to the amplitude for the coherent state, $c.c.$ is the complex conjugate. Note that the $\alpha_0$ decides which coherences will appear in the spectrum and requires special attention. For small $\alpha_0$ ($\sim$0.5), the transition from ground-state $E_0$ to first eigenstate $E_1$ dominates, while transitions from other states have little contribution. As $\alpha_0$ increases, additional transitions also start to contribute, resulting in a parametric oscillation comprised of multiple transition states. Tuning $\alpha_0$ can also induce asymmetric spectral features. 

\begin{figure}
    \centering
    \includegraphics[width=0.9\linewidth]{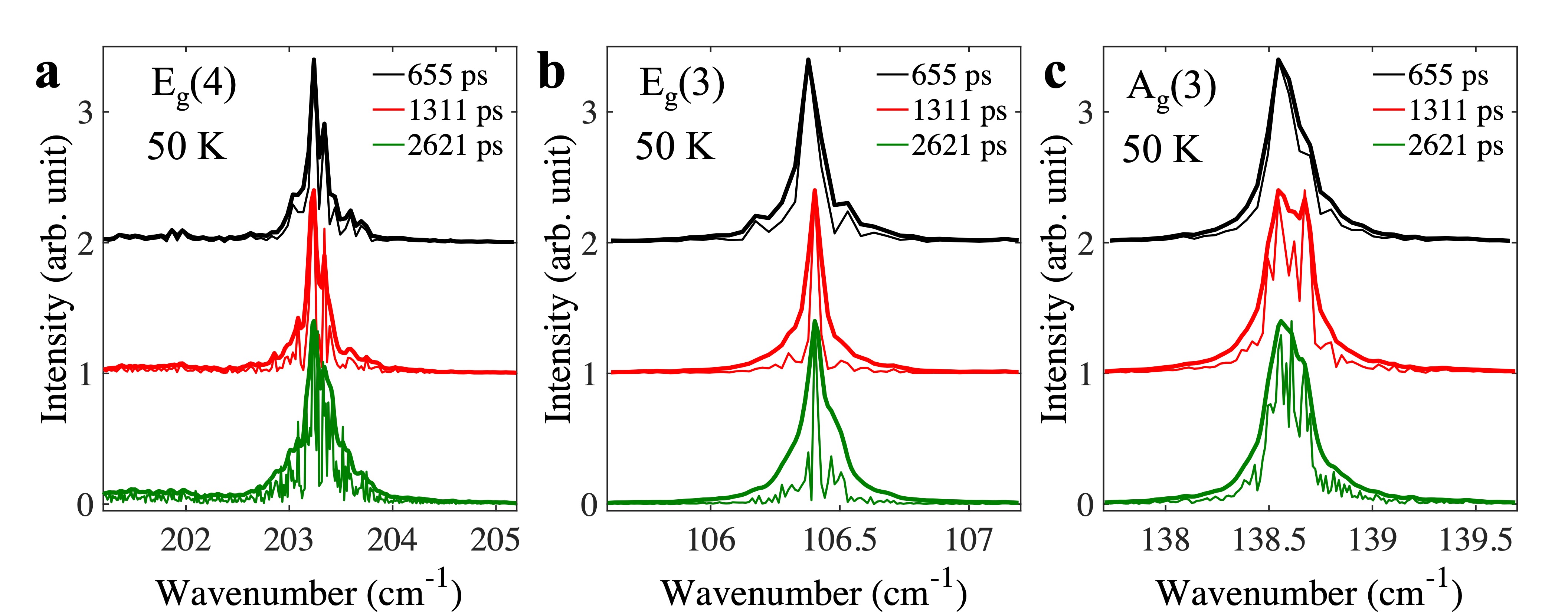}
    \caption[Egmodespacing]{(a) Spectral energy density (SED) of $E_g(4)$ mode of CrGeTe$_3$ obtained from MLMD simulations for various trajectory lengths at 50\,K. The thick lines show the spectrum convolved with a Lorentzian of FWHM 0.025\,cm$^{-1}$ for all trajectories for ease of visualization.  (b, c) Same as panel (a) but for $E_g(3)$ and $A_g(3)$ mode at 50\,K.}
    \label{fig_CGT_othermodes_AIMD}
\end{figure}

Now to calculate $\langle x \rangle_t$ for the $A_g(5)$ mode, we first reconstruct the phonon potential such that phonon eigenstates obtained from solving the Schr\"odinger equation have the same spacing as in the SED data.  The reconstructed phonon potential and corresponding phonon eigenstates at 50 and 200\,K are shown in SI Fig.~\ref{fig_CGTSEDEVPcomparison}e. These eigenstates are then subsequently used in Eq.~\eqref{expectationvalue_main} to calculate $\langle x \rangle_t$. Figure~\ref{fig_CGTphononspectra}c shows the parameteric oscillation at 50 and 200\,K comprising primarily the first eight and six eigenstates lasting $\sim$100 picoseconds. Here, we used $\omega = 289.86$ and 288.23\,cm$^{-1}$, $\alpha_0 = 1.8$ and 1.35, and a small exponential decay ($\gamma$) to account for the losses at 50 and 200\,K (see discussion on variation of parameters in SI Section~\ref{Oscillatormodel}). The spectral intensity can be calculated by Fourier transforming the obtained spectrum, i.e., $I(\omega) \propto (\langle x \rangle_\omega)^2$  (blue curve in panel (a)), which shows satellite peaks and their intensity consistent with the SED data, except for one peak at $\sim$289.24\,cm$^{-1}$, which is broadened beyond a clear signature. We further investigated modes of other symmetry at the zone-center, for example, see SI Fig.~\ref{fig_CGT_othermodes_AIMD} for the $E_g(4)$, $E_g(3)$, and $A_g(3)$ modes at 50\,K, but did not find conclusive evidence of frequency combing. From the above results, we conclude that the SED data, coupled with the reproduction of the frequency comb and its intensity using a phenomenological model, indicate a comb-like feature in the $A_g(5)$ mode, but phonon broadening from scattering hampers its clear identification. Nevertheless, since the peaks are not distinct, strictly, they do not qualify as frequency combs.  

\begin{figure}
    \centering
    \includegraphics[trim=2cm 0cm 4cm 0cm, clip=true,width=0.9\linewidth]{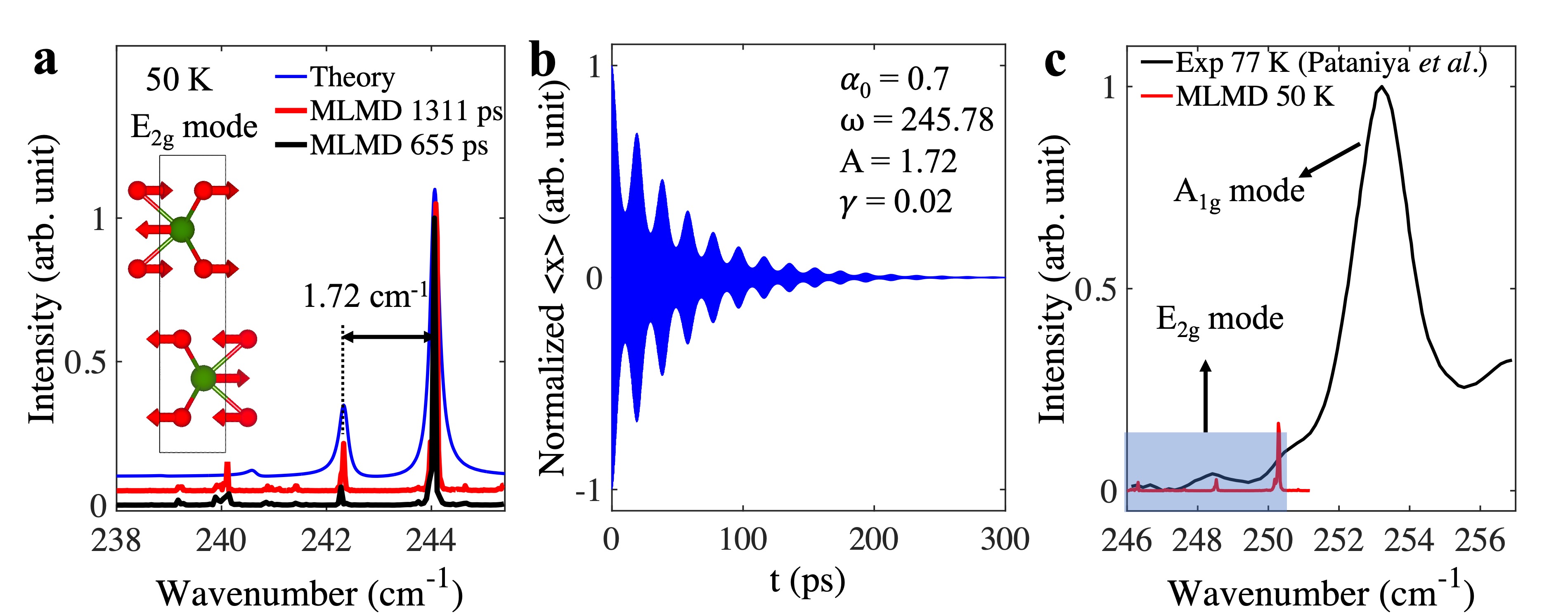}
    \caption[WSe2 $E_{2g}$ mode spacing]{(a) Frequency combs of the $E_{2g}$ mode of WSe$_2$ as obtained by calculating SED of zone-center modes from MLMD simulation at 50\,K for different trajectories, and from phenomenological model. The spectra are vertically offset for better visibility. (b) $\langle x \rangle_t$ obtained from the phenomenological model. (c) SED data obtained from MLMD simulations at 50\,K are overplotted with experimental data at 77\,K from Ref.~\citenum{pataniya2020low}. The bigger peak at 253\,cm$^{-1}$ is from the $A_{1g}$ mode. To match experimental and simulated energies, the simulated data are rigidly shifted by 6.2\,cm$^{-1}$.}
    \label{fig_WSe2_potential}
\end{figure}

\begin{figure}
    \centering
    \includegraphics[width=0.6\linewidth]{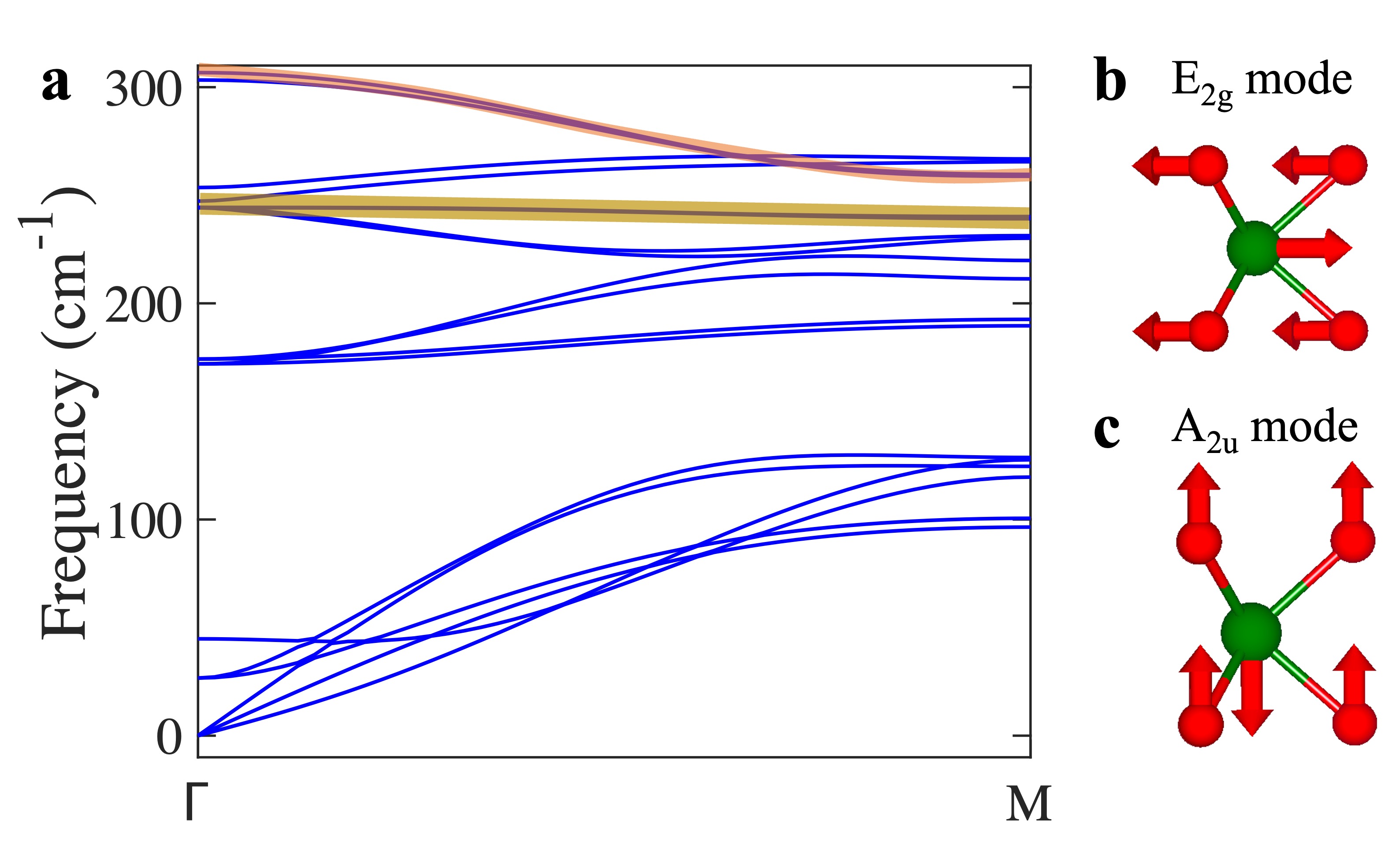}
    \caption[WSe2 Phonon dispersion and $E_{2g}$ eigenvector]{(a) Phonon dispersion of	 WSe$_2$ along $\Gamma$-M direction. The $E_{2g}$ (yellow) and $A_{2u}$ (brown) phonon mode is highlighted. The phonon eigenvectors of (b) $E_{2g}$ and (c) $A_{2u}$ modes.}
    \label{fig_WSe2phonondispersion}
\end{figure}

\begin{figure}
    \centering
    \includegraphics[width=0.8\linewidth]{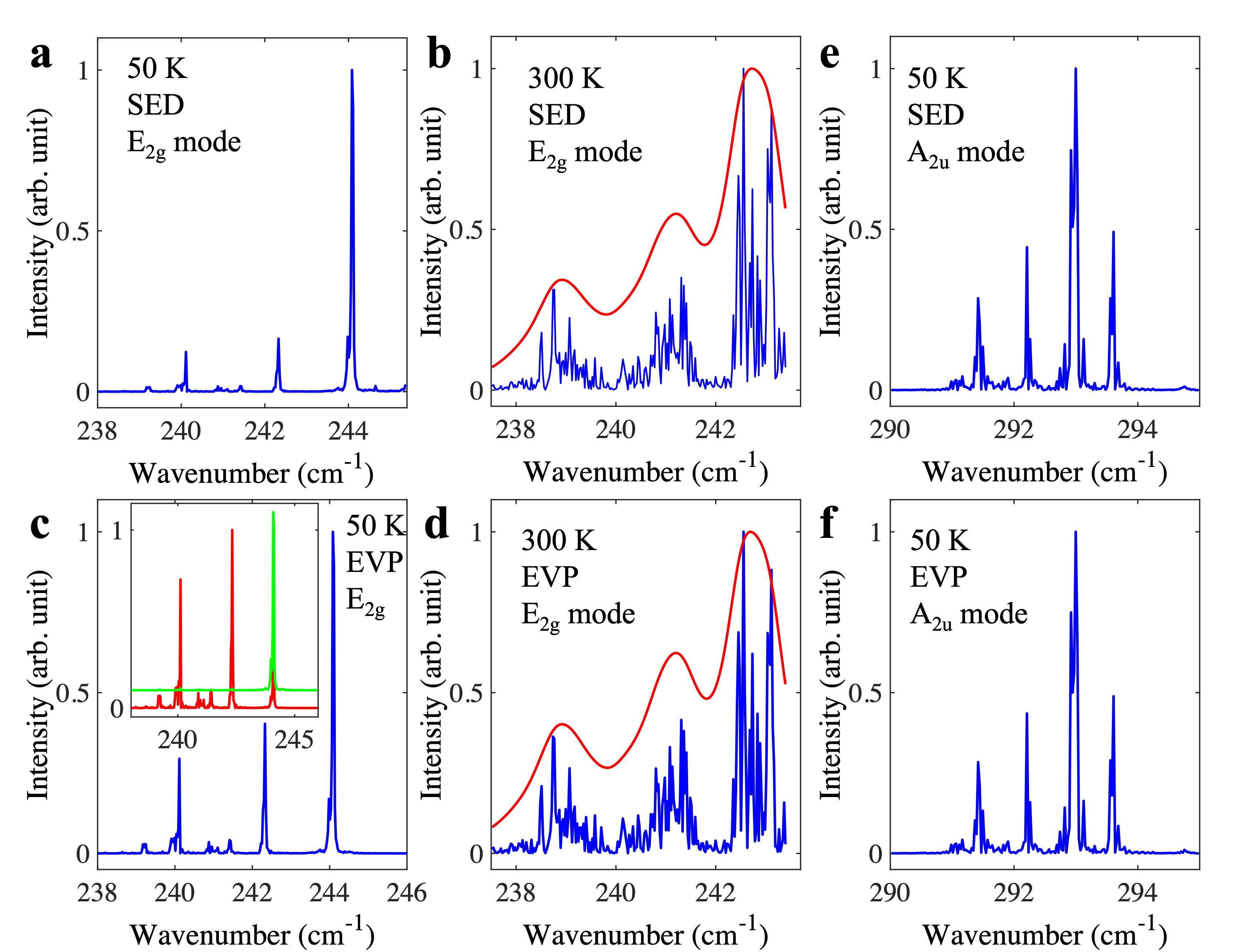}
    \caption[WSe2 SED and EVP Comparison]{(a-d) Comparison of spectral energy density (SED) and zero-Kelvin eigenvector-projected (EVP) spectrum for the $E_{2g}$ mode (sum of both degenerate modes, see inset) at 50 and 300\,K for trajectory length of 1311\,ps.  The small variation in intensity could be due to a change in the phonon eigenvectors with temperature. Inset in panel (c) shows the EVP of two degenerate $E_{2g}$ modes. Curves are vertically offset for clarity. The red curve in panels (b,d) is convolved with a Lorentzian having a FWHM of 0.4\,cm$^{-1}$ for ease of visualization. (e,f) Same as panels (a) and (c) but for the $A_{2u}$ mode.}
    \label{fig_SEDEVPcomparison}
\end{figure}

The above detailed analysis provided us with feedback on the optimal phonon potential and the minimal phonon scattering required for spontaneous frequency combing. We found that nearly flat intramolecular $E_{2g}$ and $A_{2u}$ zone-center phonon modes of WSe$_2$ (see SI Fig.~\ref{fig_WSe2phonondispersion}), a 2D van der Waals solid ($P6_{3}/mmc$ space group), exhibited large self-interaction anharmonicity with minimal phonon scattering, showing clear signatures of combs beyond the experimental resolution. The calculated value of $k_3/k_2$ for the $E_{2g}$ mode ($\sim$30.5\,meV) is $\sim$72\% and $k_4/k_2$ for the $A_{2u}$ mode ($\sim$37.1\,meV) is 26.2\% for $Q_D = 0.43$\,\AA. Note that the $A_{2u}$ mode potential is symmetric; hence, its cubic term drops out. Similar to the case of CrGeTe$_3$, SED of the $E_{2g}$ mode corresponding to two different trajectory lengths is shown in Fig.~\ref{fig_WSe2_potential}a. To ensure that all observed features originate from the same $E_{2g}$ mode, we projected the 0\,K eigenvector on the MLMD trajectory.  The projected spectral data is shown in SI Fig.~\ref{fig_SEDEVPcomparison}, and is consistent with peaks observed from SED calculations. Observation of sharp satellite peaks originating from the same phonon mode strongly suggests the spontaneous formation of a frequency comb. To understand their occurrence, we use the same phenomenological model discussed above. Figure~\ref{fig_WSe2_potential}b shows the parameteric oscillation comprising the first four eigenstates lasting 100s of picoseconds. Here, we used $\omega = 245.8$\,cm$^{-1}$ and $\alpha_0 = 0.7$. A small difference in peak position of the satellite peak near 240\,cm$^{-1}$ is due to exclusion of quartic and higher-order anharmonic terms, which will make the comb spacing increase or decrease depending on the sign, unlike for cubic anharmonicity, for which the comb spacing remains constant. A numerical approach is needed to include quartic and higher-order terms, which we use later to simulate the $A_{2u}$ mode. As described above, the $k_3/k_2$ ratio (or equivalently $\lambda/( \frac{1}{2}\mu \omega^2)$ in the normalized $x$ coordinate) for the $E_{2g}$ mode is 72\%, suggesting a strongly anharmonic potential (See SI Fig.~\ref{fig_E2g_A2u_potential}a); however, the observed peaks are sharp, thus indicating the presence of limited phonon scattering channels. As discussed earlier, the presence of both phonon self-interaction anharmonicity and a limited scattering channel in the $E_{2g}$ mode allows the spontaneous frequency comb generation in the $E_{2g}$ mode.

Since the comb spacing observed here is $\sim$1.7\,cm$^{-1}$, it should be visible in high-resolution Raman scattering measurements on high-quality crystals. We find preliminary evidence in the literature for the frequency comb. Figure~\ref{fig_WSe2_potential}c overplots our SED data at 50\,K, shown in panel (a), with Raman data measured at 77\,K for vapor-phase synthesis-grown crystals.~\cite{pataniya2020low} Note that in the unpolarized Raman measurements, both $E_{2g}$ and $A_{1g}$ phonons appear at nearby frequencies, making them challenging to separate, which we address using Raman tensor simulation (see SI Section\ref{Ramantensorcalculation}) and previous helicity-resolved data.~\cite{chen2015helicity} From the Raman simulation and previous data, we concluded that the small peaks correspond to the $E_{2g}$ mode in measurements (highlighted in light blue in panel (c)). These peaks compare reasonably well with frequency comb simulations. Further, we compare our simulations at 300\,K with reported data on a 4-layer WSe$_2$ sample (see SI Fig.~\ref{fig_E2g_A2u_potential}c). Here, the helical polarization of incident and scattered light allows only the $E_{2g}$ mode to appear in the measured configuration.~\cite{chen2015helicity} The measured spectrum, with a primary peak at 250\,cm$^{-1}$ and unusually broadened features between 240 and 250\,cm$^{-1}$, is consistent with the simulated SED data. We must emphasize that the multiple peaks at 77\,K and the asymmetric broad features at 300\,K observed in the measurements are not due to phonon-phonon interactions; if they were, our SED data would have captured them. Instead, our SED data show an intensity profile consistent with that of a frequency comb. We caution that other Raman studies on WSe$_2$ do not show as clear a feature as Refs.~\citenum {pataniya2020low} and~\citenum{chen2015helicity}, possibly due to instrument resolution or sample quality, but careful observation does reveal broad features. Nevertheless, this comparison with measurements should only be treated as preliminary and not as conclusive evidence. Moreover, in CrGeTe$_3$, isotopes were found to be the origin of the comb-like features in measurements for the $A_g(5)$ mode (see SI Section~\ref{CGTisotopes}).~\cite{krasucki2025spin} Therefore, we explicitly exclude any role of W or Se isotopes that could account for the various combs observed for the $E_{2g}$ mode in Raman measurements (see SI section~\ref{WSe2isotopes}). We further rule out the possibility of hot bands, as the intensity of the adjacent peak will be three orders of magnitude smaller than that of the fundamental peak at 50\,K, and will be within the baseline of the Raman spectrum (see SI Section~\ref{sectionhotband}).

\begin{figure}
    \centering
    \includegraphics[trim=0cm 0cm 0cm 0cm, clip=true,width=0.75\linewidth]{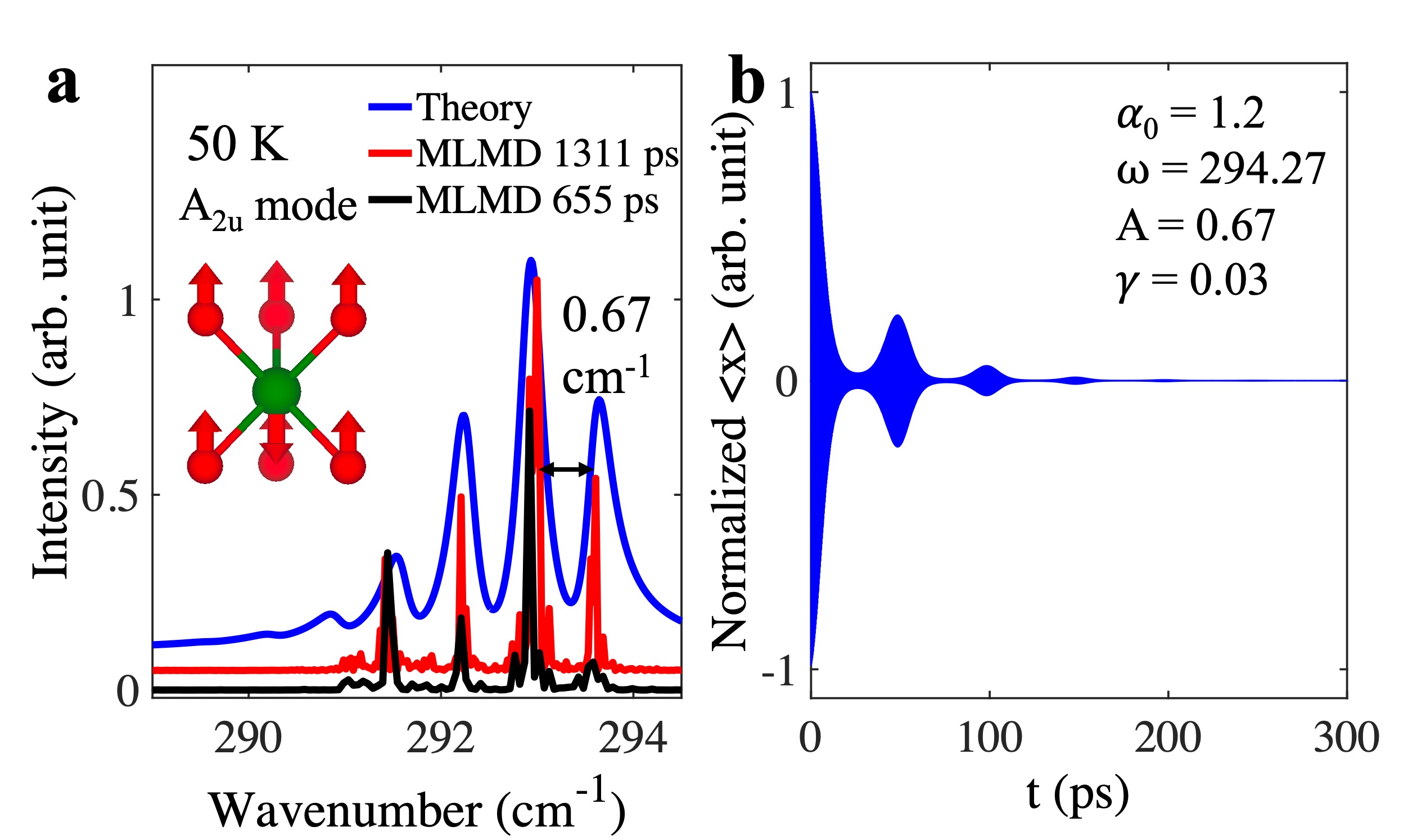}
    \caption[WSe2 $A_{u}$ mode spacing]{(a,b) Same as Fig.~\ref{fig_WSe2_potential}(a,b) but for the $A_{2u}$ mode of WSe$_2$.}
    \label{fig_WSe2_Aumode_potential}
\end{figure}

Next, we repeat the same analysis for the $A_{2u}$ mode, an infrared-active intramolecular out-of-plane phonon mode of WSe$_2$ (see eigenvector in inset of Fig.~\ref{fig_WSe2_Aumode_potential}a). The spontaneous formation of a frequency comb is evident from SED data and persists across different trajectory lengths. We further projected the 0\,K eigenvector of the $A_{2u}$ mode on the MLMD trajectory leading to the same spectral feature (see SI Fig.~\ref{fig_SEDEVPcomparison}), consistent with peaks observed from SED calculations. Because of the symmetry of the phonon mode, here we take the fourth-order anharmonic term (the third-order term will be zero) and calculate the phonon eigenstates by numerically solving the Schrödinger equation, such that the comb spacing matches the SED data. Note that we do not use relative intensity information from SED data to match the comb spacing. The same numerically calculated energy spacings are then used to calculate $\langle x \rangle_t$, which shows the parameteric oscillation comprising primarily the first five eigenstates lasting 100s of picoseconds (see Fig.~\ref{fig_WSe2_Aumode_potential}b). Here, we used $\omega = 294.3$\,cm$^{-1}$ and $\alpha_0 = 1.2$. Subsequently, the Fourier transform of the trajectory reproduces the frequency comb, including the relative intensity of various combs, as shown by the blue color curve in panel (a). Notably, the phonon potential here is anharmonic with a $k_4/k_2$ ratio of 26.2\%. The presence of a strong anharmonic potential (see SI Fig.~\ref{fig_E2g_A2u_potential}b), together with the limited phonon scattering channels, as evidenced by the sharp peaks, fulfills the conditions for spontaneous frequency comb generation. For the $A_{2u}$ mode, we did not find infrared data in the literature to compare with simulations. However, we note that the isotope contribution would broaden the $A_{2u}$ mode by $\sim$4\,cm$^{-1}$ (see SI Fig.~\ref{fig_WSe2isotopes}b), thereby making frequency combs experimentally difficult to decipher in the absence of isotopically pure crystals.

In summary, we provide a quantitative understanding of the spontaneous formation of frequency combs. Our MLMD simulations reveal comb-like features in the $A_g(5)$ mode of CrGeTe$_3$ at 50 and 200\,K; however, the comb spacings are comparable to $\Gamma$, making precise identification challenging. Based on a comprehensive analysis of CrGeTe$_3$, we identified the conditions for spontaneous frequency comb generation -- a strong anharmonicity governed by self-interaction and limited phonon scattering. The above conditions are met for two modes of WSe$_2$, which show a clear signature of frequency combing. Our results provide guidance for engineering long-lived and spectrally resolved phonon states for applications in precision sensing, signal processing, and hybrid quantum technologies. 

%

\clearpage
\newpage

\clearpage
\newpage

\section*{Supplementary Information}

\subsection{Ab initio and Machine-Learned Molecular Dynamics} \label{MLMDsimulationdetails}
\begin{figure}[h]
    \centering
    \includegraphics[width=0.4\linewidth]{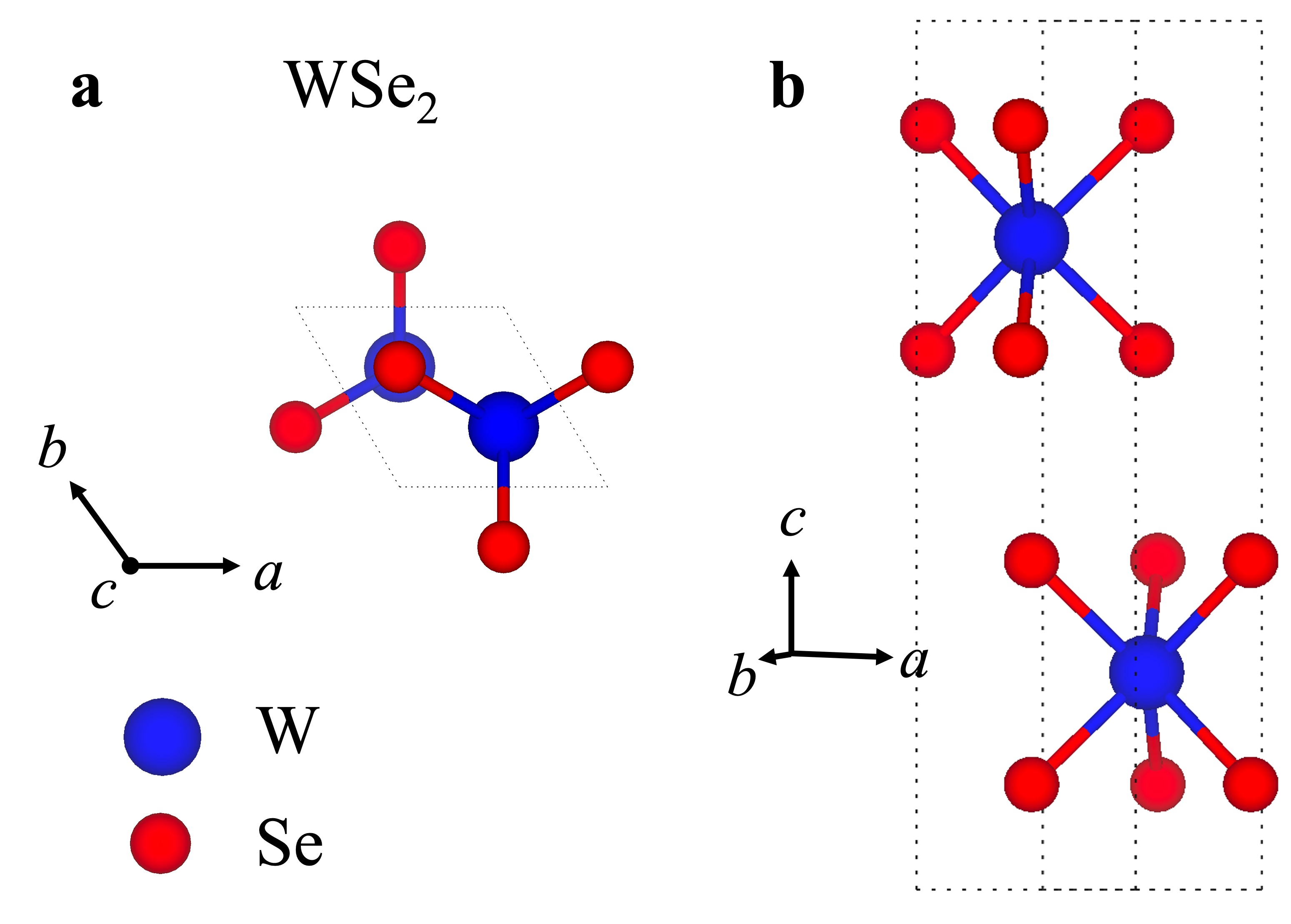}
    \caption[WSe2-crystalstructure]{(a,b) Crystal structure of WSe$_2$ from different views. Blue and red denote W and Se atoms.}
    \label{fig_WSe2_crystalstructure}
\end{figure}

\begin{figure}
    \centering
    \includegraphics[width=0.9\linewidth]{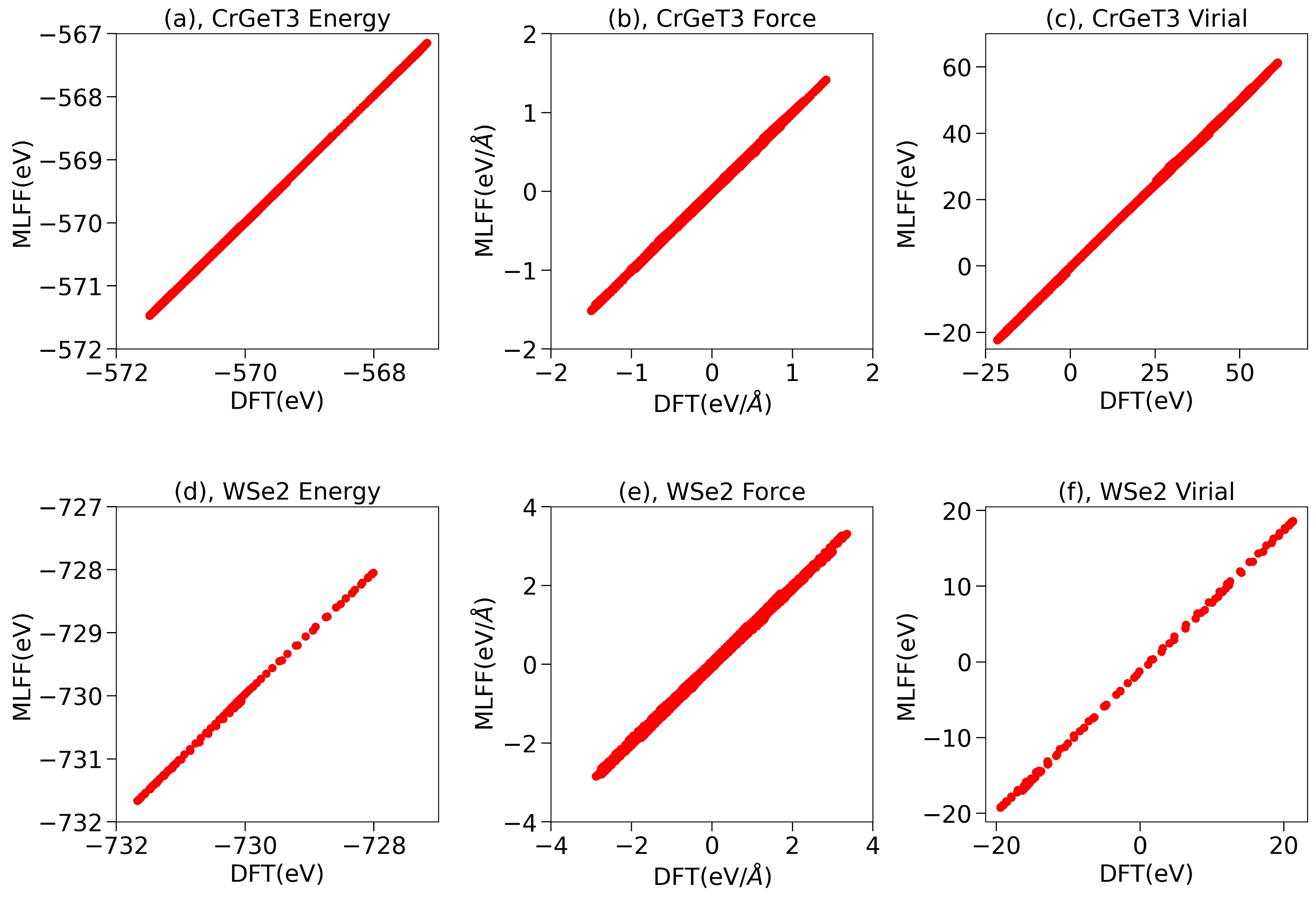}
    \caption[CGT MLFF validation]{The \emph{ab initio} molecular dynamics (AIMD) and machine learned molecular dynamics (MLMD) calculated total energies, forces, and Virial for (a-c) CrGeTe$_3$ and (d-f) WSe$_2$. Here, the $x$ axis shows the energy calculated from DFT, and the $y$ axis shows the energy calculated using the machine-learned force field (MLFF). The good agreement between the two simulations validates the machine-learned force field.}
    \label{fig_CGTMLFFvalidation}
\end{figure}

Ab initio molecular dynamics (AIMD) simulations were carried out within the density functional theory (DFT) framework as implemented in the Vienna Ab initio Simulation Package (VASP).~\cite{kresse1996efficient,kresse1996efficiency,kresse1993ab} The projector-augmented wave (PAW) formalism was employed together with the generalized gradient approximation (GGA) for the exchange–correlation functional, using the Perdew-Burke-Ernzerhof (PBE) parametrization. Van der Waals interactions between WSe$_2$ layers along the c-axis were accounted for using the DFT-D3 method. At ambient conditions, WSe$_2$ crystallizes in a hexagonal structure (space group P63/mmc, No. 194, see Fig.~\ref{fig_WSe2_crystalstructure}) containing six atoms per unit cell (two W and four Se atoms). A 4$\times$ 4$\times$ 1 supercell comprising 96 atoms was used for the AIMD simulations. DFT calculations employed a 2 $\times$2$\times$2 Monkhorst–Pack k-point mesh and a plane-wave energy cutoff of 800\,eV. The total energy convergence criterion was set to 10$^{-6}$\, eV. AIMD simulations were performed in the temperature range of 100–800\,K, in steps of 100\,K, within the canonical (NVT) ensemble. Temperature control was achieved using a Nosé–Hoover thermostat. For CrGeTe$_3$, AIMD simulations were conducted using a 2$\times$2$\times$1 supercell of the conventional unit cell (space group $R\overline{3}$, No. 148). A plane-wave cutoff energy of 600\,eV and a 3$\times$3$\times$4 k-point mesh were used for Brillouin zone integration. On-site Hubbard corrections for Cr 3$d$ electrons were included within the Dudarev approach, with an effective Hubbard parameter $U_{\rm{eff}}$ = 3.5\,eV.~\cite{wang2023magnetic} AIMD simulations for CrGeTe$_3$ were performed at temperatures of 10, 50, 100, 300, and 500\,K. Harmonic phonon calculations were carried out using the finite-displacement method as implemented in the open-source Phonopy package, employing the same computational parameters as used in the AIMD simulations. Machine-learned surrogate force fields for WSe$_2$ and CrGeTe$_3$ were generated using the DeePMD framework,~\cite{wang2018deepmd} based on a neural network representation. Comprehensive AIMD datasets were used to train the neural network potentials. A cutoff radius of 8.0\,\AA\,was chosen for atomic interactions, while the embedding and fitting network sizes were set to (25, 50, 100) and (240, 240, 240), respectively.
Machine-learned molecular dynamics (MLMD) simulations were performed using the trained DeePMD potentials within the LAMMPS package.~\cite{plimpton1995fast} The generated machine-learned force field was tested against several quantities computed from AIMD (see Fig.~\ref{fig_CGTMLFFvalidation}). Phonon spectral energy density (SED) calculations were conducted based on MLMD trajectories using a 10$\times$10$\times$10 supercell, corresponding to 6,000 atoms for WSe$_2$. For CrGeTe$_3$, a 10$\times$10$\times$3 supercell having 9,000 atoms is used for the hexagonal unit cell having 30 atoms. Simulations were run for up to $\sim$2.6\,ns, yielding a momentum and energy resolution of 0.1 reciprocal lattice units and approximately 0.001\,meV ($\sim$0.01\,cm$^{-1}$), respectively.

\begin{figure}
    \centering
    \includegraphics[width=0.8\linewidth]{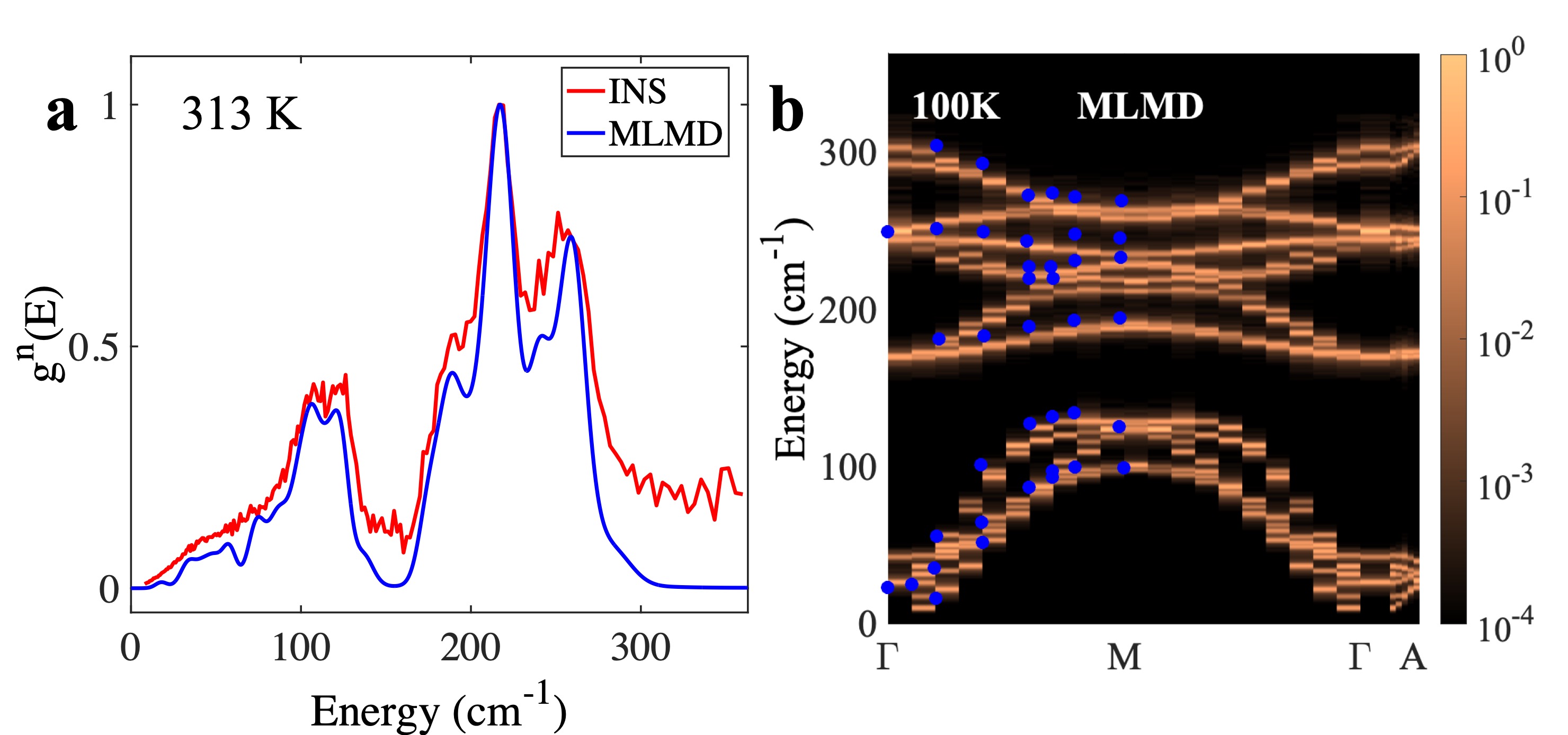}
    \caption[WSe2 phonon dos and dispersion comparison]{(a) Comparison of phonon DOS obtained from MLMD simulations and previously reported neutron-weighted inelastic scattering measurements at 313\,K.~\cite{gupta2023distinct} (b) Comparison of phonon spectral energy density calculated from MLMD simulations and measurements (blue marker) at 300\,K by Cai \it{et al}.~\cite{cai2022monolayer}}
    \label{fig_WSe2_dos_dispersioncomparison}
\end{figure}

The phonon spectral energy density at wavevector $\vec{q}$ and energy $E$ can be defined as:~\cite{thomas2010predicting}
\begin{equation}
\varphi(\vec{q}, E) = 
\frac{1}{4\pi \tau_0 N} 
\sum_{\alpha,k} m_k\left| 
\sum_{n=1}^N \int_0^{\tau_0} 
\vec{\dot{u}}_{\alpha}\!\left(
\begin{matrix}
n\\
k
\end{matrix}
; t
\right)
\exp\left[ i \vec{q}. \vec{r}
\begin{pmatrix}
n \\
k
\end{pmatrix}
-\frac{i E t}{\hbar}  \right] \, dt \right|^2,
\end{equation}
where $N$ is the number of unit-cells in a supercell $(N = N_1\times N_2\times N_3)$, and the summation index $\alpha$ runs over Cartesian axes, while index $k$ runs over the number of particles in the unit cell. We note $m_k$, the mass and $\vec{r}\begin{pmatrix}
n \\
k
\end{pmatrix} $, the equilibrium position of the $k^{th}$ atom in the $n^{th}$ unit cell, and $\vec{\dot{u}}_{\alpha}\!\left(
\begin{matrix}
n\\
k
\end{matrix}
; t
\right)$, its velocity at time $t$. A molecular dynamics simulation with a supercell dimension $(N_1\times N_2\times N_3)$ and trajectory length of $\tau_0$\,(in ps) gives a momentum and energy resolution of $\Delta \vec{q} = \frac{2 \pi}{a N_1} \hat{i} + \frac{2 \pi}{a N_2} \hat{j}+ \frac{2 \pi}{a N_3} \hat{k} $ and $\Delta E$=4.136/$\tau_0$\,meV, respectively. Here, $a$ is the lattice parameter of the conventional cubic cell. The calculated phonon spectral energy density and phonon density of states from MLMD of WSe$_2$ are overplotted with experiment data, showing excellent agreement (see Fig.~\ref{fig_WSe2_dos_dispersioncomparison}).

The eigenvector projected phonon spectral energy density of mode $\nu$ at wavevector $\vec{q}$ and energy $E$ can be defined as:~\cite{thomas2010predicting}
\begin{align}
\varphi_{\nu}(\vec{q}, E)
&= \frac{1}{4\pi \tau_0 N}
\left|
\int_{0}^{\tau_0}
\sum_{n=1}^{N} \sum_{k}
\sqrt{m_k}\,
\vec{\dot{u}}\!\left(
\begin{matrix}
n\\
k
\end{matrix}
;\, t
\right)
\cdot
{e}^{\,*}(\vec{q}, \nu, n, k)\,
\exp\!\left[
i \vec{q} \cdot
\vec{r}
\begin{pmatrix}
n\\
k
\end{pmatrix}
- \frac{iEt}{\hbar}
\right]
\, dt
\right|^2.
\end{align}
We can rewrite the expression as:
\begin{align}
\varphi_{\nu}(\vec{q}, E)
&= \frac{1}{4\pi \tau_0 N}
\left|
\int_{0}^{\tau_0}
\xi(\vec{q}, \nu, t)\,
\exp\!\left(-\frac{iEt}{\hbar}\right)
\, dt
\right|^2 ,
\end{align}
where, 
\begin{align}
\xi(\vec{q}, \nu, t)
&=
\sum_{n=1}^{N} \sum_{k}
\sqrt{m_k}\,
\vec{\dot{u}}\!\left(
\begin{matrix}
n\\
k
\end{matrix}
;\, t
\right)
{e}^{\,*}(\vec{q}, \nu, n, k)\,
\exp\!\left[
i \vec{q}\cdot\vec{r}
\begin{pmatrix}
n\\
k
\end{pmatrix}
\right].
\end{align}
Here, ${e}^{\,*}(\vec{q}, \nu, n, k)$ is the phonon mode eigenvector of $k^{th}$ atom in the $n^{th}$ unit-cell of mode $\nu$ at wavevector $\vec{q}$. All the above analyses were performed using the MDLAB code.\cite{gupta2025molecular} 

\begin{figure}
    \centering
    \includegraphics[width=0.8\linewidth]{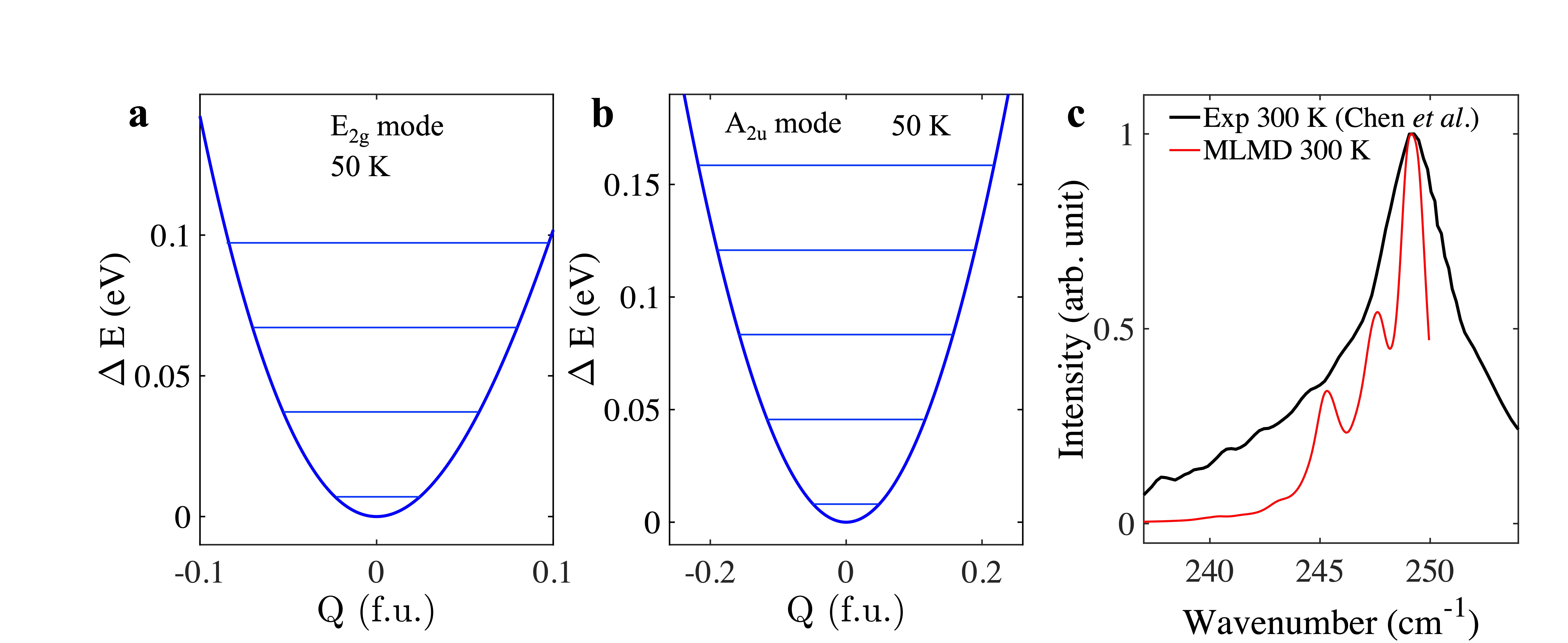}
    \caption[WSe2 $E_{2g}$ and $A_{2u}$ mode potentials]{(a) The cubic anharmonic potential of the $E_{2g}$ mode of WSe$_2$ at 50\,K, corresponding to a frequency comb with a spacing of 1.7\,cm$^{-1}$.  (b) Quartic potential of $A_{2u}$ mode of WSe$_2$ at 50\,K, corresponding to a frequency comb with a spacing of 0.67\,cm$^{-1}$. Here 1\,f.u. corresponds to 0.43\,\AA\,for both modes. (c) SED data obtained from MLMD simulations at 300\,K are overplotted with experimental data at 300\,K from Ref.~\citenum{chen2015helicity}. Due to the chosen helicity, only $E_{2g}$ mode was visible. To match the experimental and simulated energies, the simulated data are rigidly shifted by 6.4\,cm$^{-1}$. The red curve is convolved with a Lorentzian with FWHM $=0.4$\,cm$^{-1}$ to smooth the SED data for visualization.}
    \label{fig_E2g_A2u_potential}
\end{figure}

\clearpage

\subsection{Derivation of classical motion of atoms within a phonon mode} \label{Oscillatormodel}
We calculated the temporal evolution of the oscillations following the derivation given by Chen \emph{et al.} in Ref.~\citenum{chen2025spontaneously}, which we reproduce below for completeness. As frequency combs require nonlinearity, cubic or higher-order terms are required for their generation. We introduce a small, cubic nonlinear term in the harmonic oscillator Hamiltonian:
\begin{equation}
H = \frac{p^2}{2\mu} + \frac{1}{2}\mu \omega^2 x^2 + \lambda x^3
\end{equation}
Here, $p$ is the momentum of the mode, $\mu$ is the reduced mass of the oscillator, and $\lambda$ is the cubic anharmonicity.  The energy difference between adjacent eigenstates is calculated by applying second-order perturbation theory to a harmonic oscillator Hamiltonian with a small cubic nonlinear term ($\lambda x^3$):
\begin{equation} \label{equation_enegydifference}
E_{n+1} - E_n = \hbar \omega - \hbar A (n+1),
\end{equation}
where $A$ is positive and equal to the spacing between the adjacent peaks, $\omega$ is the frequency of the mode. Frequency combs are achieved by a superposition of these states, for which a semi-classical approach is used. Using oscillations within the phonon mode as quantum mechanical, and the interaction between the oscillations and the laser as classical, the complex eigenvalue for the laser-excited coherent state is given as:
\begin{equation}
|\alpha_0\rangle = e^{-\frac{|\alpha|^2}{2}} \sum_{n=0}^\infty \frac{\alpha_0^n}{\sqrt{n!}} |n\rangle.
\end{equation}
The time evolution of the expected values is calculated from the coherent state by introducing a phase factor $e^{-\iota E_n t/\hbar}$ for each eigenstate. 
\begin{equation}\label{annihilation}
\langle x \rangle_t = \langle a^+ + a \rangle_t = e^{-|\alpha|^2} \sum_{n=0}^\infty \frac{|\alpha_0|^{2n}\alpha_0}{n!} e^{-\iota (E_{n+1} - E_n)t/\hbar} + c.c.
\end{equation}
Now, using Eqn.~\ref{equation_enegydifference} in Eqn.~\ref{annihilation}, we obtain
\begin{equation} 
\langle x \rangle_t = e^{-|\alpha|^2} \sum_{n=0}^\infty \frac{|\alpha_0|^{2n}\alpha_0}{n!} e^{-\iota (\omega - A)t}e^{\iota A n t} + c.c.
\end{equation}
Rearranging the terms leads to 
\begin{equation}
\langle x \rangle_t = \alpha_0 e^{-|\alpha|^2} e^{-\iota (\omega - A)t} \sum_{n=0}^\infty \frac{(|\alpha_0|^{2} e^{\iota A t})^n}{n!}  + c.c.
\end{equation}
The summation to infinity can be rewritten as an exponential, i.e., $\sum_n^\infty \frac{z^n}{n!} = e^z$, where $z = |\alpha_0|^2e^{\iota At}$, which simplifies the expectation value of $x$ describing the classical motion of atoms within the phonon mode to 
\begin{equation} \label{expectationvalue}
\langle x \rangle_t = \alpha_0 \mathrm{exp}[-\iota (\omega-A)t + |\alpha_0|^2(e^{\iota A t}-1)]+c.c.
\end{equation}
We also include exponential decay ($\gamma$) in the expectation value to account for the phonon-phonon and other scattering terms that lead to losses. The Fourier transformation of Eqn.~\ref{expectationvalue} maps the classical time-domain motion of the phonon mode into the frequency domain, where phononic frequency combs can be directly seen.

\begin{figure}
    \centering
    \includegraphics[width=0.8\linewidth]{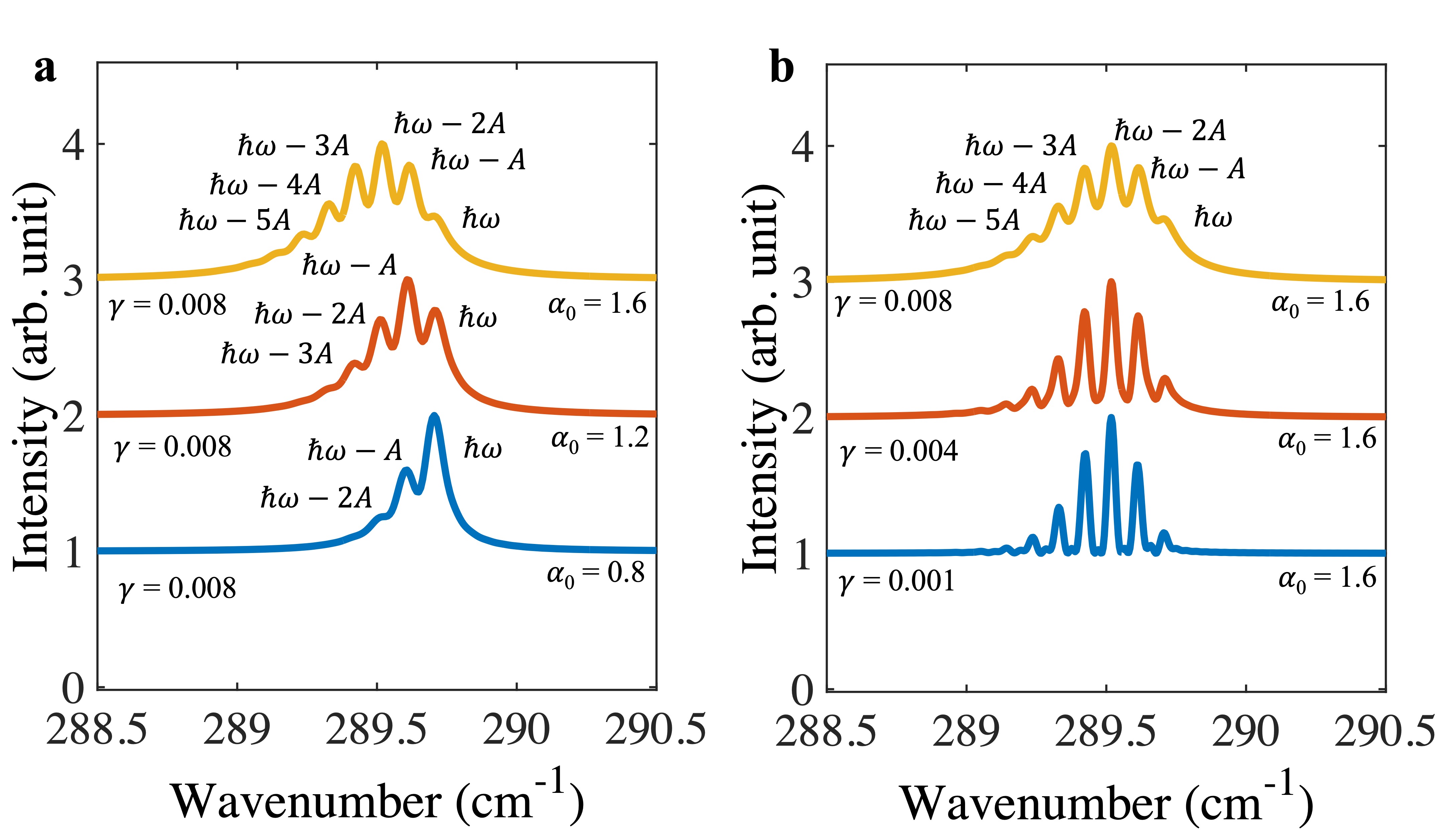}
    \caption[Variation of $\alpha_0$ and $\gamma$]{a) Comb-like features from phenomenological model for different $\alpha_0$. $\alpha_0$ denotes a complex eigenvalue for the coherent state, and acts as a window that controls the number of eigenstates participating in the coherence. All curves are here with $\gamma = 0.008$. (b) Comb-like features from phenomenological model for different $\gamma$ with $\alpha_0 =1.6$. The $\gamma$ does not change the relative intensities or spacings of the peaks.}
    \label{suppfig28_alphagammavariation}
\end{figure}

To understand the role of various parameters in the phenomenological model, we first show the effect of varying $\alpha_0$, which represents a complex eigenvalue of the coherent state. Smaller values of $\alpha_0$ correspond to coherent superpositions of fewer phonon eigenstates, while larger values correspond to superpositions of more eigenstates. The coherent state $\langle x \rangle$ and the relative intensities of the frequency comb are outputs of the phenomenological model. For a given value of $\alpha_0$, the model produces a unique set of relative intensities across all peaks (see Fig.~\ref{suppfig28_alphagammavariation}a); the relative intensity of individual combs cannot be independently tuned.

Next, we focus on the decay factor $\gamma$. Varying $\gamma$ increases or decreases the losses in the oscillator and controls the peak broadening as shown in Fig.~\ref{suppfig28_alphagammavariation}b. However, $\gamma$ variation does not affect the relative intensities or positions of the peaks. At very large damping values, excessive broadening will cause individual peaks to lose coherence and become indistinguishable as discrete spectral features.

\begin{figure}
    \centering
    \includegraphics[width=0.7\linewidth]{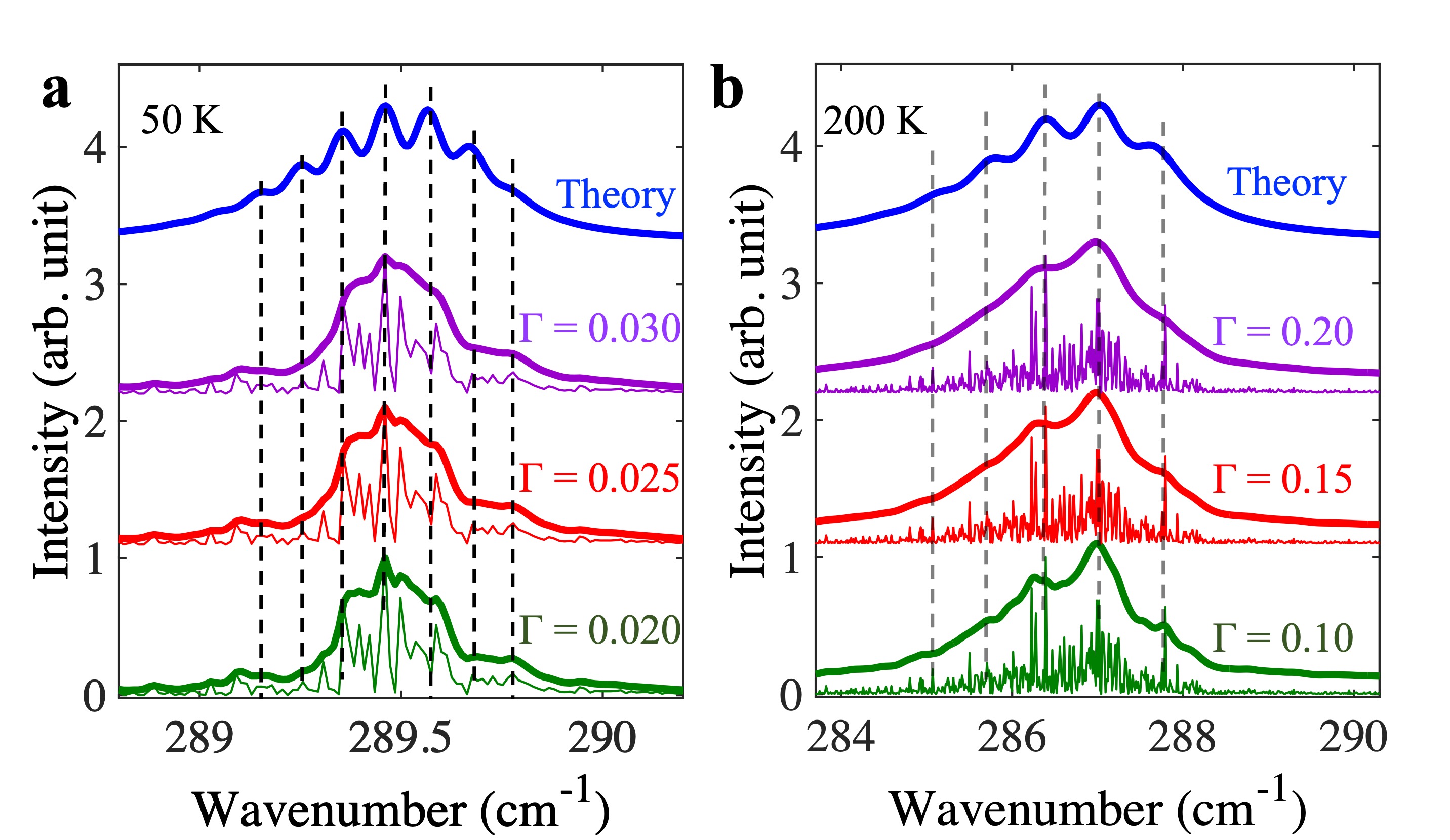}
    \caption[Variation of $\Gamma$ at 50 and 200\,K]{(a) Comparison between various $\Gamma$ of Lorentzian convolutions (0.020, 0.025, and 0.030\,cm$^{-1}$) for CrGeTe$_3$ at 50\,K to smooth the SED data. The trajectory length is 2621\,ps. The peaks are consistently captured for $\Gamma$ values. Raw data are shown as thin lines, and smoothened (convolved) spectra are shown as thick lines. (b) Same as panel (a) but for CrGeTe$_3$ at 200\,K.}
    \label{suppfig29_differentbroadening_200K}
\end{figure}

We should mention here that when comparing the simulated raw SED data with the phenomenological model output, we use a Lorentzian convolution to smooth the raw SED data for visualization. However, this smoothing has no physical meaning. If we had infinite trajectory length from MLMD simulations to calculate SED, there would have been no need to smooth the curve with a Lorentzian function for visualization. We compare results for different FWHM values, i.e., $\Gamma$ at 50 and 200\,K, and find that the comb-like feature is captured reasonably well across all FWHM values, indicating that the comb-like features are independent of the Lorentzian broadening FWHM within this limit (see Fig.~\ref{suppfig29_differentbroadening_200K}). A large FWHM will broaden the peaks, making them merge into a single feature. 

\clearpage
\subsection{Spacing from phonon potential} \label{spacingcalculation}
The energy spacing between the phonon states is calculated by numerically solving the time-independent Schr\"odinger wave equation for a given phonon potential $V$ in MATLAB,
\begin{equation}
H\psi = -\frac{\hbar^2}{2\mu} \frac{\partial^2 \psi}{\partial x^2}+V\psi = E \psi.
\end{equation}
Here, $x$ is the phonon mode displacement, and $\mu$ is the reduced mass associated with the phonon mode.

\subsection{Distinction between anharmonicity from phonon self-interaction and phonon-phonon interaction} \label{anharmonicity}

As discussed in the main text, the spontaneous formation of phonon frequency combs requires the presence of anharmonicity. Here, we distinguish between the anharmonicity arising from phonon self-interaction and phonon-phonon interaction. As explained below, the former does not lead to decay of oscillation amplitude, but the latter does. 

\begin{figure}
    \centering
    \includegraphics[width=0.8\linewidth]{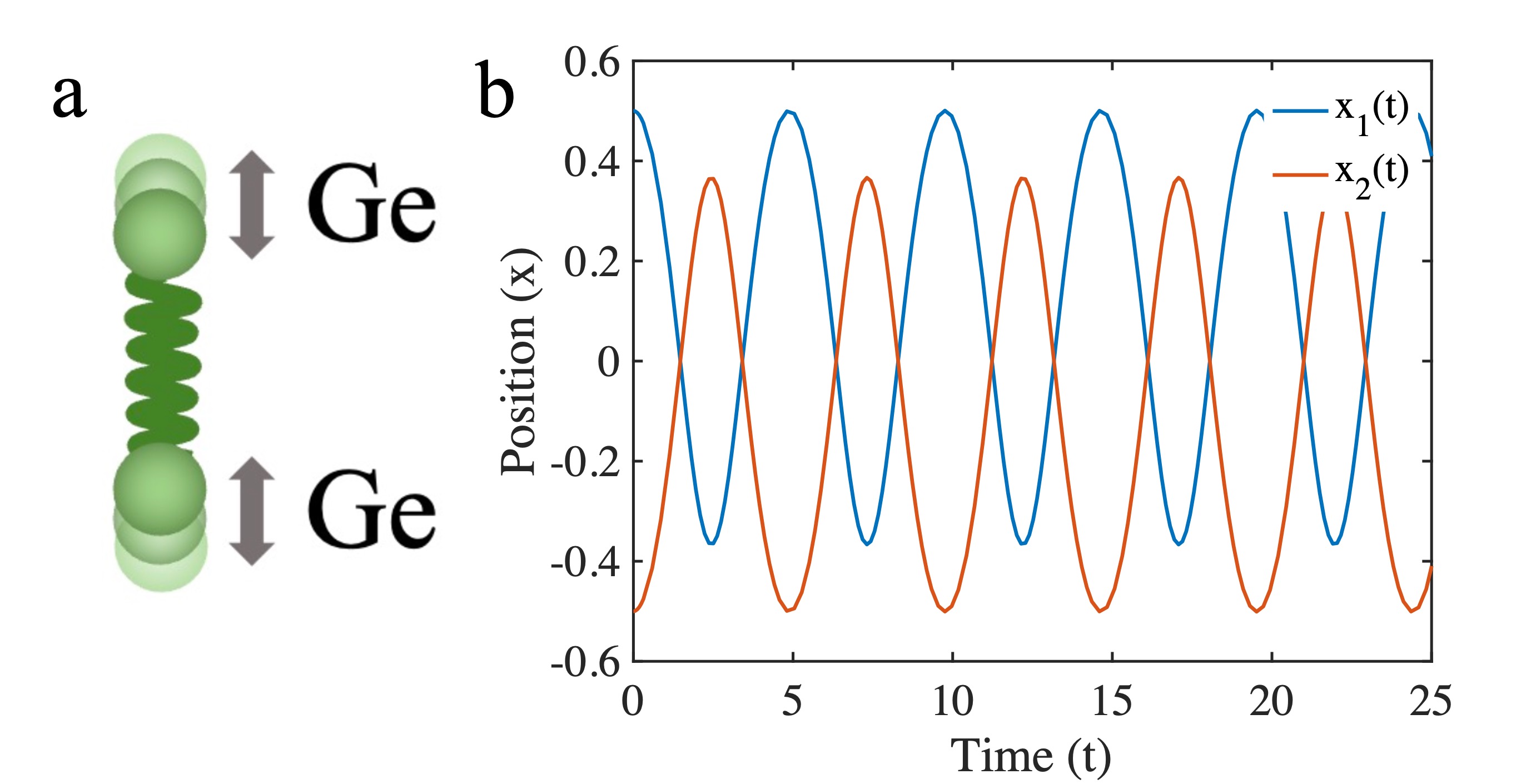}
    \caption[Spring mass model]{a) Schematic of the $A_g(5)$ mode, showing the Ge-Ge dimer. (b) The two masses connected by a nonlinear spring will oscillate indefinitely with time in the absence of damping.}
    \label{suppfig26_springmassmodel}
\end{figure}

Note that the anharmonic potential itself does not necessarily imply a finite lifetime; only in the case where anharmonicity arises from system-system interactions (exchange of energy/momentum between two systems) does the system amplitude decay in time. We take an example of two masses connected by a spring. The spring force can be linear with displacement or non-linear. In the former, the potential will be harmonic, and in the latter, it will be anharmonic. Let the nonlinear spring force be given by:
\begin{equation}
F = -k(x_2-x_1)-\alpha(x_2-x_1)^2.
\end{equation}
Here, $k$ is the linear spring constant, $\alpha$ is the nonlinear coefficient, $x_1$ and $x_2$ are the displacements of two masses from equilibrium. The equation of motion of two masses can be written as, 
\begin{equation}
m_1 \ddot{x}_1 = k(x_2-x_1)+\alpha(x_2-x_1)^2.
\end{equation}
\begin{equation}
m_2 \ddot{x}_2 = -k(x_2-x_1)-\alpha(x_2-x_1)^2.
\end{equation}
Solving this differential equation numerically by taking $k=1,\alpha=1/2,m_1=m_2=1$, both masses oscillate indefinitely without decay in the absence of damping (see Fig.~\ref{suppfig26_springmassmodel}b). The potential can be calculated by integrating the spring force, and the ratio of the cubic to the quadratic term in the potential is 67\%. Thus, it is evident from the spring-mass system that, if an anharmonic potential arises from self-interaction (here, the spring force between two masses), the system can sustain oscillations indefinitely. 

Now we focus on the spacing of the phonon frequency combs. As shown schematically in the main text Fig.~\ref{fig_figure1}d and discussed in the main manuscript, solving the Schr\"odinger equation for the harmonic potential yields equally spaced energy levels and results in a single delta function in the frequency domain. But in the case of a cubic potential, energy levels are not equally spaced; rather, their spacing decreases by $A$ for each higher energy level ($A_1$ = $A$, $A_2$ = $2A$,  $A_3$ = $3A$, ... and so on, see main text Fig.~\ref{fig_figure1}e). These unequally spaced energy levels give rise to multiple delta functions in energy space, each separated by a spacing $A$. However, if we take an analogy with the spring-mass system interacting with its surroundings, meaning it exchanges energy and momentum with its surroundings, then the oscillatory amplitude will decay, the delta function will become Lorentzian with a finite FWHM $\Gamma$, smearing the discrete features, as shown in the main text Fig.~\ref{fig_figure1}f. This smearing of discrete features is exactly what happens if the anharmonicity arises from phonon-phonon interactions. Phonon-phonon interactions induce the finite frequency shift and lead to a finite phonon linewidth $\Gamma$, which can be determined by calculating the phonon self-energy $\Sigma = \Delta +\iota\Gamma$,~\cite{Bruesch} and is detrimental to observation of phonon frequency combs. The Ge-Ge dimer vibration of the $A_g(5)$ mode in CrGeTe$_3$ is one such example where the vibration is not fully isolated and interacts with other phonons, as evidenced by calculations of the phonon-phonon scattering phase space (see main text Fig.~\ref{fig_CGTphononspectra}d), leading to a finite $\Gamma$ comparable to the spacing between the comb-like features of the $A_g(5)$ mode, and thus masking them. However, if the system is nearly isolated with intramolecular anharmonicity (same as spring-mass system discussed above), as is the case with the $E_{2g}$ and $A_{2u}$ modes of WSe$_2$, which is evidenced by the negligible broadening of the phonon peaks in the MLMD simulations, the combs are clearly visible.

\clearpage

\subsection{Scattering phase space} \label{Scattering_phase_space}
Three-phonon scattering phase space with the occupation factor is calculated as, 
\begin{equation}
W^{(\pm)}(\textbf{q}j) = \frac{1}{N_\textbf{q}} \sum_{\textbf{q}_1,\textbf{q}_2,j_1,j_2} {\begin{Bmatrix}n_2-n_1 \\ n_1+n_2+1 \end{Bmatrix}}\delta(\omega_{\textbf{q}j} \pm \omega_{\textbf{q}_1j_1}-\omega_{\textbf{q}_2j_2}) \delta_{\textbf{q} \pm \textbf{q}_1,\textbf{q}_2+\textbf{G}}.
\end{equation}
Here, \textbf{q} is the phonon wavevector, $n_1 = n(\omega_{\textbf{q}_1 j_1})$, $n_2 = n(\omega_{\textbf{q}_2 j_2})$, $n$ is the occupation factor given by $n(\omega) = \frac{1}{e^{\hbar \omega/k_\textrm{B} T - 1}}$ and is temperature dependent, $\omega$ is the phonon frequency, and $j$ is the mode number.~\cite{tadano2014anharmonic}

\begin{figure}
    \centering
    \includegraphics[width=0.8\linewidth]{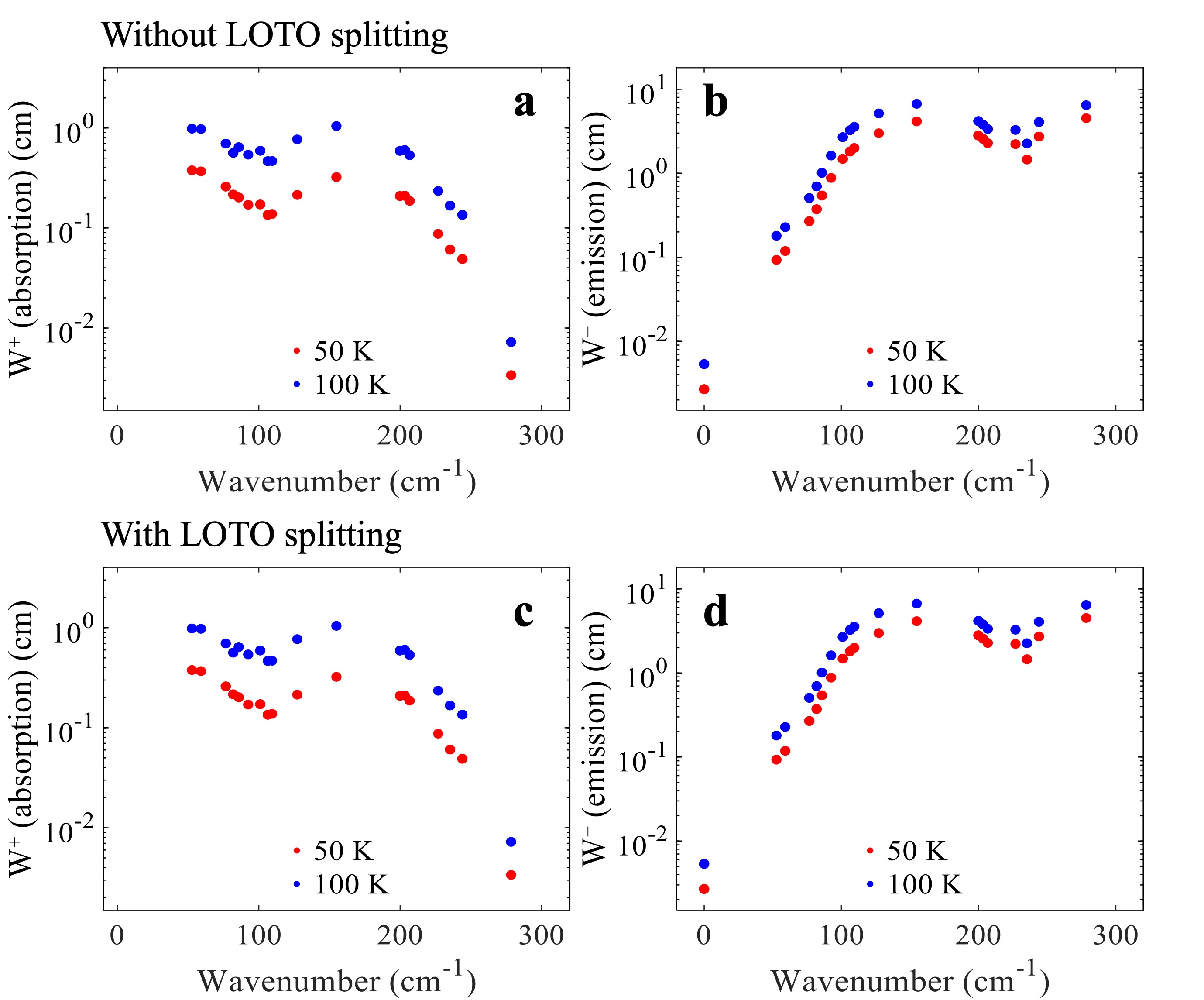}
    \caption[LOTOsplitting]{(a,b) Phonon scattering phase space without and (c,d) with LO-TO splitting at 50\,K and 100\,K. Minimal change is observed in the absorption or emission scattering phase space due to the LO-TO splitting.}
    \label{fig_LOTOsplitting}
\end{figure}

We calculated the $W$ for CrGeTe$_3$, and find that while $W^+$ for the $A_g(5)$ mode is negligible, $W^-$ is finite even at 50\,K, which further increases with temperature (main text Fig.~\ref{fig_CGTphononspectra}d). We further re-evaluated $W$ without considering the LO-TO splitting, but it did not make an appreciable difference in the scattering phase space, thus ruling out any long-range Coulomb interaction on the scattering phase space (see Fig.~\ref{fig_LOTOsplitting}).
\clearpage
\subsection{Spin-phonon coupling} \label{spinphononcoupling}

\subsubsection{Unperturbed Exchange interaction} \label{unperturbedexchangeinteraction}
\begin{figure}
    \centering
    \includegraphics[width=0.9\linewidth]{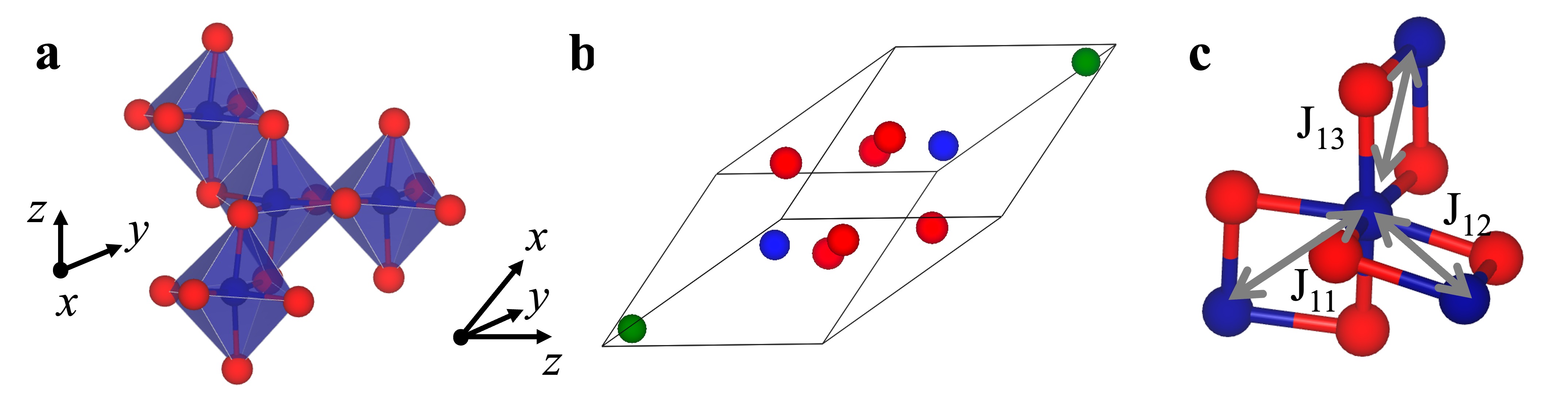}
    \caption[Primitivecell]{(a)  Crystal structure of CrGeTe$_3$ showing the NN Cr atoms bonded via edge-sharing CrTe$_6$ polyhedra. (b) The primitive cell of CrGeTe$_3$ contains 10 atoms, comprising two of each Cr and Ge atom, and six Te atoms. Red, green, and blue colors denote Te, Ge, and Cr atoms, respectively. (c) The labeling of nearest neighbour (NN) exchange interactions $J_{11}$, $J_{12}$, and $J_{13}$ in CrGeTe$_3$.}
    \label{suppfig1:primitivecell}
\end{figure}

\begin{figure}
    \centering
    \includegraphics[width=0.7\linewidth]{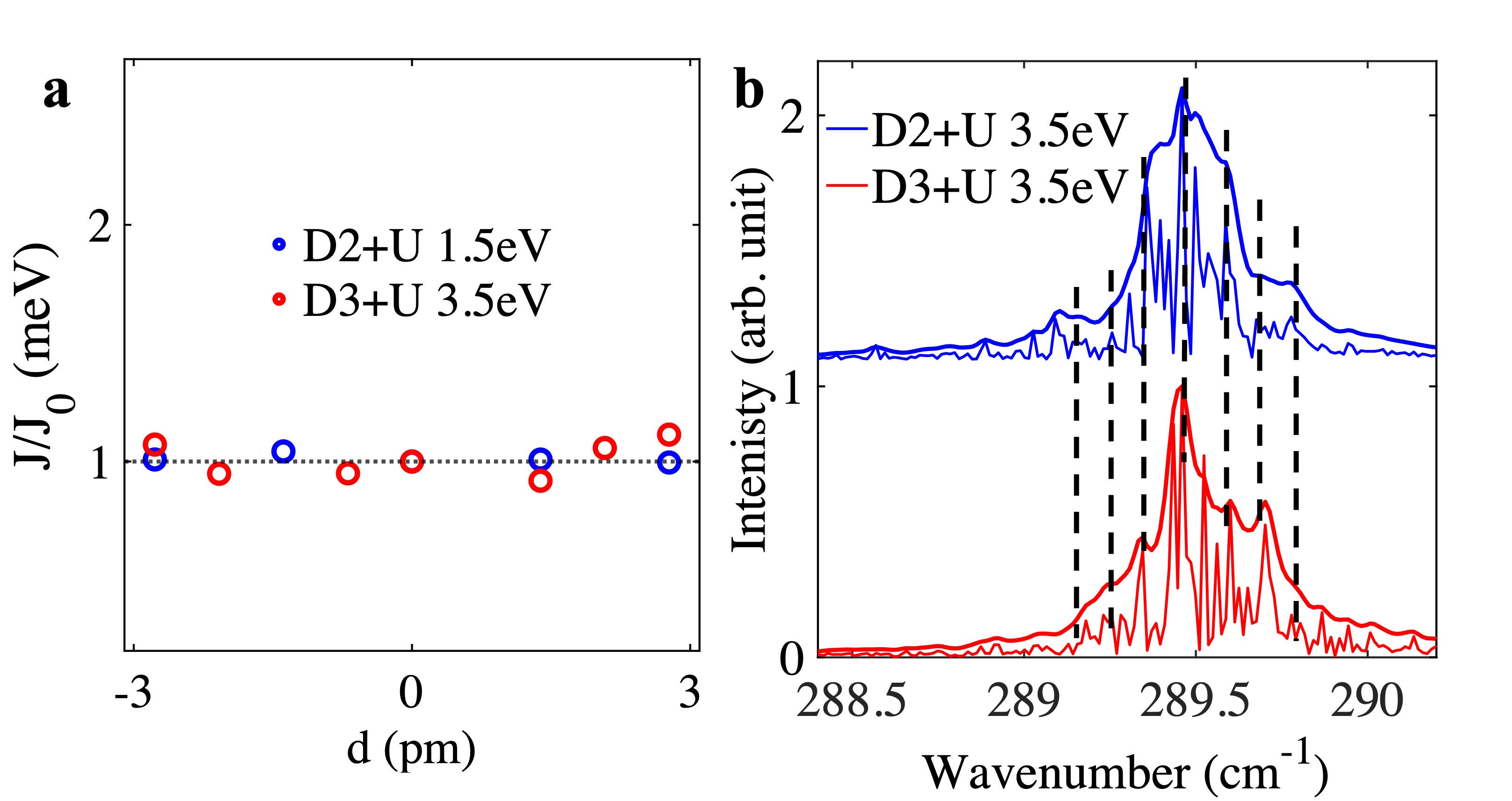}
    \caption[CrGeTe$_3$ D2 D3 comparison]{(a) Variation of exchange interaction with phonon perturbation for D2+$U$ 1.5\,eV and D3+$U$ 3.5\,eV configurations. There is little to no variation in $J$s ($J = (J_{11}+J_{12}+J_{13})/3$) for perturbation by the eigenvectors of the $A_g(5)$ mode, which is indicative of negligible spin phonon coupling in the mode. (b) A comparison of SED of $A_g(5)$ mode obtained from MLMD simulations using the D2+$U$ 3.5\,eV and D3+$U$ 3.5\,eV vdW method showing nearly identical response.}
    \label{fig_D2_D3}
\end{figure}

CrGeTe$_3$ crystallizes in a rhombohedral $R\overline{3}$ space group (No. 148) with slightly distorted octahedra. CrGeTe$_3$ undergoes a transition into a ferromagnetic (FM) state with magnetic moments aligned along the $c$-axis below the Curie temperature of approximately 65\,K.~\cite{carteaux1995crystallographic} We performed spin-polarized electronic structure calculations in the DFT framework as implemented in Quantum Espresso,~\cite{QE-2017, QE-2009} on a primitive cell with 10 atoms - six Te and two each of Cr and Ge atoms (see Fig.~\ref{suppfig1:primitivecell}). The plane-wave cutoff energy is set to 120\,Ry for converged results. The lattice parameters of the primitive cell are $a=b=c=7.615$\,\AA\, and $\alpha = \beta = \gamma = 51.495^\circ$. We used a $4 \times 4 \times 4$ $\Gamma$-centered Monkhorst-Pack electronic $k$-point mesh to integrate over the entire Brillouin zone. The energy change is less than $10^{-3}$\,eV/atom on increasing mesh size from $4 \times 4 \times 4$ to $6 \times 6 \times 6$. The structure is optimized by relaxing atomic positions until forces on all atoms are less than 1\,meV per atom. We used a generalized gradient approximation (GGA) in the Perdew-Burke-Ernzerhof (PBE)~\cite{refPBE} parametrization with a Hubbard correction. To account for the localized Cr $d$ orbital electrons, GGA+U calculations are used,~\cite{Dudarev_1998} with onsite Coulomb interaction $U_{\rm{eff}}$ varying between 1.5 and 3.5\,eV.~\cite{fang2018large} For the van der Waals dispersion interactions, we adopt both the DFT-D2 and DFT-D3 methods. Since the primary vibration under consideration here involves Ge-Ge dimers rather than interlayer interactions, the results are nearly identical for the DFT-D2 and DFT-D3 vdW methods (see Fig.~\ref{fig_D2_D3}). The upper and lower bounds of the band gap for different values of $U_{\rm{eff}}$ and vdW methods range from $\sim$0.26 to 0.37\,eV, comparable to the experimentally observed band gap.~\cite{li2018electronic} Zone center phonons were calculated using density functional perturbation theory (DFPT) as implemented in Quantum Espresso, and are in agreement with the reported phonon modes using Raman scattering.~\cite{krasucki2025spin} There are five $A_g$ modes (88, 132, 145, 241, and 294\,cm$^{-1}$) and five $E_g$ modes (80, 95, 116, 233, and 250\,cm$^{-1}$) at the zone center.

\begin{figure}
    \centering
    \includegraphics[width=1\linewidth]{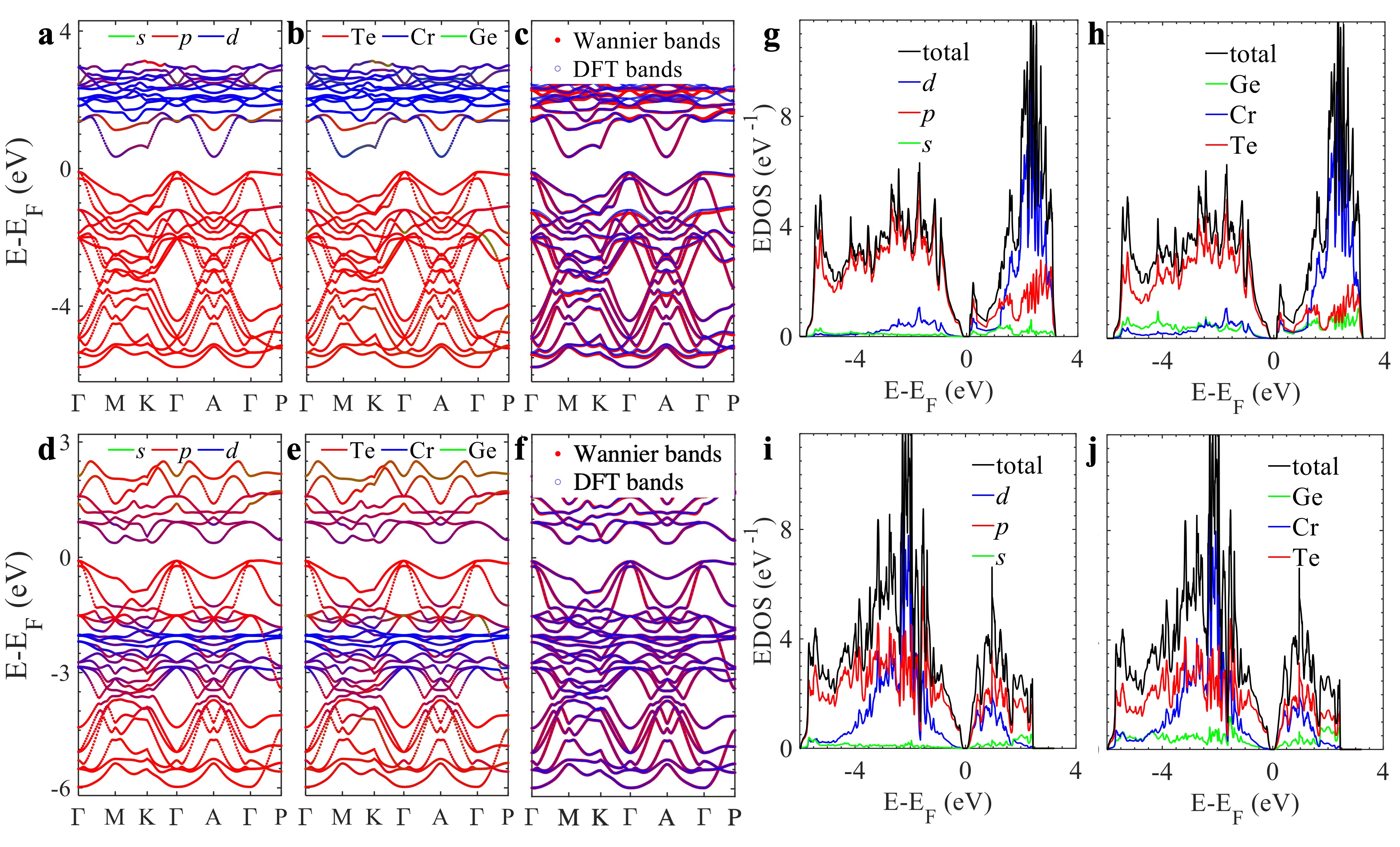}
    \caption[Band structure of CrGeTe$_3$]{(a,b) Orbital- and atom-resolved band structure of CrGeTe$_3$ along high symmetry directions for the down-spin configuration. (c) Comparison between band structures from DFT and Wannier projections, confirming the accuracy of the Wannier projections for the down-spin configuration. Plots d-f are the same as a-c, except for the up-spin configurations. (g,h) Orbital- and atom-resolved electronic density of state (EDOS) of CrGeTe$_3$ for the down-spin configuration. (i,j) Same as panels (g,h) but for the up-spin configuration.}
    \label{fig_bandstructure}
\end{figure}

\clearpage

To visualize the orbital orientations, we computed the maximally localized Wannier projections on a 30-orbital basis by enforcing the maximum localization~\cite{pizzi2020wannier90} on 10 $d$ orbitals of two Cr atoms, 2 $s$ orbitals of two Ge atoms, and 18 $p$ orbitals of six Te atoms to cover the entire bandwidth of $d$ orbitals of Cr and $p$ orbitals of Te atoms. The electronic band structure obtained from Wannier projections matches the DFT-calculated band structure (see Fig.~\ref{fig_bandstructure}c,f). We also computed the maximally localized Wannier projection by constraining the Wannier functions' symmetry to be the same as pure atomic $d$ orbitals.~\cite{korotin2015calculation} Following the methodology of Ref.~\citenum{korotin2015calculation}, the one-electron Hamiltonian matrix obtained by Wannier function projection on constraining the symmetry to pure $d$ orbitals is used to calculate the intersite Green’s function at every {\bf k} point in the reciprocal space. The intersite Green's function between any two atoms is calculated by integrating Green's function over the entire Brillouin Zone (BZ). The resulting intersite Green's function is used in the analytic expression for the exchange integrals:
	\begin{equation}\label{Jcalculationequation}
		J_{ij} = \frac{1}{2\pi}\int_{-\infty}^{E_\mathrm{F}} d\epsilon \sum_{mm'm''m'''} \mathrm{Im} (\Delta_{i}^{mm'}G_{ij,\downarrow}^{m'm''}\Delta_{j}^{m''m'''}G_{ji,\uparrow}^{m'''m} ),
	\end{equation}
	where $E_\textrm{F}$ is the Fermi energy, $i$ and $j$ are site indexes in the primitive cell,  $m$ and $m'$ numerate orbitals on $i$ and $j$  sites, respectively. $G_{ji,\uparrow}^{mm'}~(G_{ij,\downarrow}^{mm'})$  is the real-space intersite Green's function for spin-up~(-down) obtained by BZ integration, and 
	\begin{equation}\label{Deltaequation}
		\Delta_{i}^{mm'} = \int_\mathrm{{BZ}} [H_{ii,\uparrow}^{mm'}(\mathrm{\textbf{k}}) - H_{ii,\downarrow}^{mm'}(\mathrm{\textbf{k}})] d\mathrm{\textbf{k}}.
	\end{equation}
	Here, $H_{ii,\uparrow}^{mm'}~(H_{ii,\downarrow}^{mm'})$ is the matrix elements of the one-electron Hamiltonian for the up~(down) in the reciprocal space. We further extend the above methodology to include phonon perturbations by displacing the atoms following the phonon eigenvector and recalculating Eqs.~\eqref{Jcalculationequation} and~\eqref{Deltaequation} in the phonon-perturbed configuration (here $E_\mathrm{F}$ and $G$ are also for the phonon-perturbed configuration), which allows us to determine the orbital-resolved spin-phonon coupling. Wannierization of the phonon-perturbed configuration used the same 30-orbital basis.

\begin{figure}
    \centering
    \includegraphics[width=0.8\linewidth]{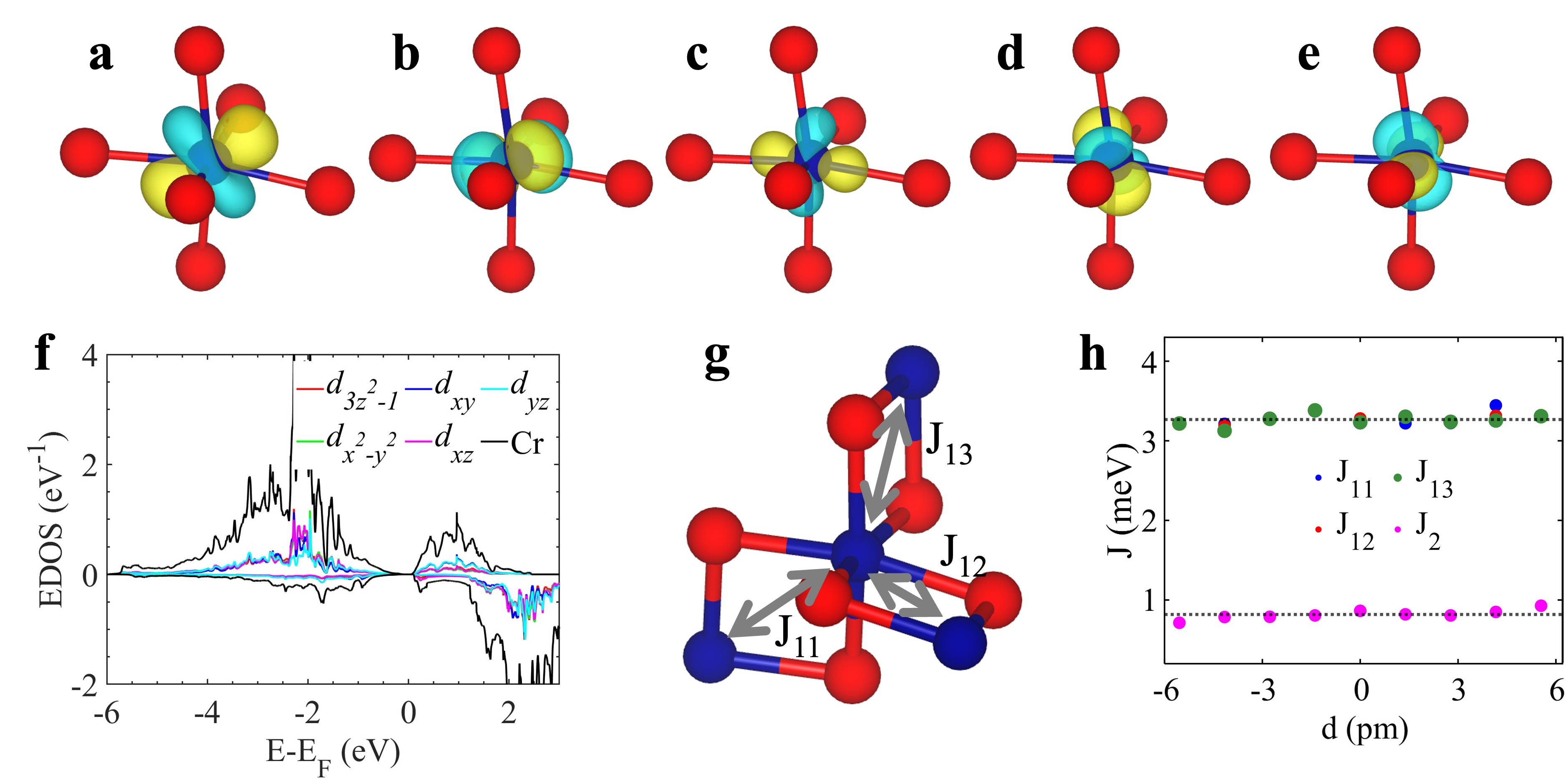}
    \caption[CrGeTe$_3$ phonon magnon scattering]{(a-e) Cr $d$ orbitals in the order $d_{3z^2-1},d_{xz}, d_{yz}, d_{x^2-y^2}$, and $d_{xy}$, respectively, as obtained from the Wannier projections. The orbitals are not oriented toward the Cr-Te bonds, resulting in partial occupancies of all Cr $d$ orbitals. (f) EDOS of Cr $d$ orbitals showing partial occupancies of all Cr $d$ orbitals. (g) The labeling of nearest neighbour (NN) exchange interactions $J_{11}$, $J_{12}$, and $J_{13}$ in CrGeTe$_3$. (h) Variation of NN exchange interactions $J_{11}$, $J_{12}$, $J_{13}$, and next NN exchange interaction $J_2$ on perturbation with $A_g(5)$ mode eigenvector.}
    \label{fig_phononmagnonscattering}
\end{figure}

The electronic density of state (DOS) of Cr $d$ orbital is given in Fig.~\ref{fig_phononmagnonscattering}f. All the orbitals are occupied compared to the typical half-filled $e_g$ orbitals for three unpaired electrons. The Cr $d$ orbitals orientations are given in Fig.~\ref{fig_phononmagnonscattering}(a-e). The $d_{3z^2-1}$ orbital is not aligned towards any Cr-O bond (same for $d_{x^2-y^2}$), whereas for an ideal octahedra, the $d_{3z^2-1}$ and $d_{x^2-y^2}$ orbitals align towards out-of-plane and in-plane ligand atoms, as observed in NiO.~\cite{ss2025effect} This asks for a detailed analysis of the orbital orientation and corresponding orbital bonding and exchange interactions in CrGeTe$_3$. 

In the unperturbed configuration, consider the exchange interaction $J_{11}$. All the isotropic orbital interactions contribute to the $J_{11}$, as all orbitals are partially occupied. For example, $J_{11}$ is the sum of $J_{3z^2-1}$, $J_{xz}$, $J_{yz}$, $J_{x^2-y^2}$ and $J_{xy}$, unlike in NiO, where only $J_{3z^2-1}$ contribute to the NNN exchange interaction, as rest of the terms are zero.~\cite{korotin2015calculation} In $J_{11}$,  the contribution from $J_{3z^2-1}$, $J_{xz}$, $J_{yz}$, $J_{x^2-y^2}$, and $J_{xy}$  are -1.51, 0.97, -0.92, -1.02, and -0.72\,meV respectively. In this convention, AFM interactions are positive, while FM interactions are negative. For $J_{12}$,  the contributions from $J_{3z^2-1}$, $J_{xz}$, $J_{yz}$, $J_{x^2-y^2}$, and $J_{xy}$  are -0.07, 0.64, -0.59, -1.56, and -1.70\,meV respectively, and for $J_{13}$, the contributions are -3.83, -1.97, 0.0, 0.82, and 1.74\,meV, respectively. Note that the three nearest neighbor (NN) $J$s -- $J_{11}$, $J_{12}$, and $J_{13}$ are degenerate in magnitude due to the same interatomic distance and bond angle; however, the orbital contributions differ as different orbitals are involved in the exchange interaction. The NN $J$ value of -3.2\,meV (ferromagnetic) is in close agreement with the value reported in the literature.~\cite{chen2022anisotropic, zhu2021topological} 

\clearpage

\subsection{Frequency comb-like features from isotopes} 
\subsubsection{CrGeTe$_3$ and CrSiTe$_3$} \label{CGTisotopes}
\begin{figure}
    \centering
    \includegraphics[width=0.7\linewidth]{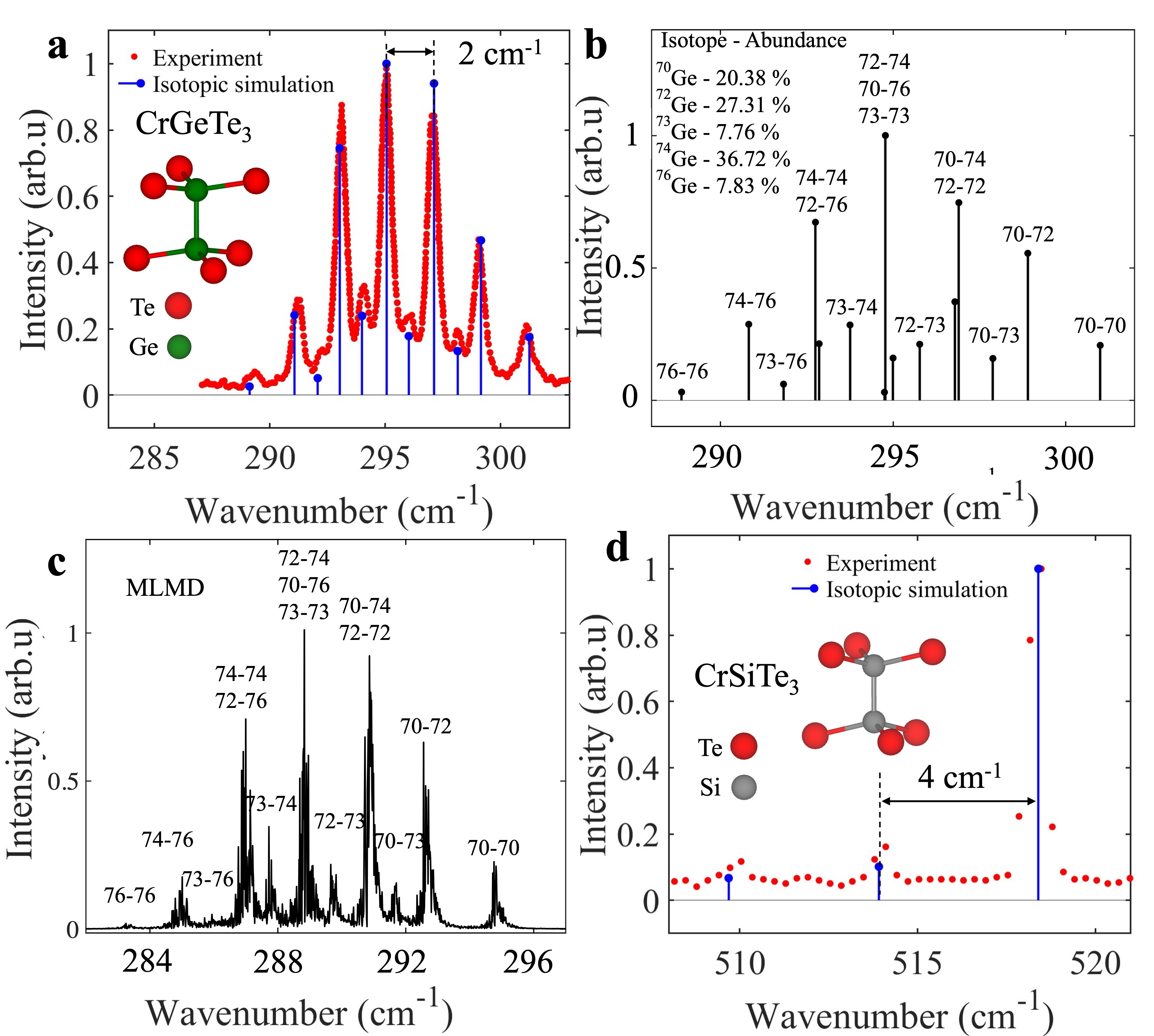}
    \caption[CrGeTe$_3$ isotope]{Phonon frequencies calculated by considering isotopes of (a) Ge in CrGeTe$_3$ and (d) Si in CrSiTe$_3$. The intensities are calculated based on the relative natural abundances. For CrGeTe$_3$, cumulative intensity is plotted, as different combinations of isotope contribute to the same frequency, for example, a combination of $^{74}$Ge$-^{74}$Ge and $^{72}$Ge$-^{76}$Ge gives the same frequency of $\sim$292.5\,cm$^{-1}$ as shown in panel (b). (c) SED of the $A_g(5)$ mode for different isotope masses as obtained from MLMD simulations. Experimental data for CrSiTe$_3$ is from Chen \emph{et al}.~\cite{chen2025spontaneously} and CrGeTe$_3$  is from Krusacki \emph{et al}.~\cite{krasucki2025spin}}
    \label{fig_CGT_Geisotopes}
\end{figure}

The $A_g(5)$ mode corresponds to a collective vibration of two neighboring Ge atoms moving towards each other, with minimal displacement of the surrounding atoms. These localized vibrations are approximated as a Ge-Ge pseudo-molecule, allowing the frequency spectrum to be described by a simple spring-mass model with two point masses connected by a spring. The phonon frequency is calculated using the linear chain model, which is widely used in shear and breathing interlayer vibration modes in van der Waals materials.~\cite{zhao2013interlayer,grzeszczyk2016raman} Naturally occurring Ge isotopes include $^{70}$Ge, $^{72}$Ge, $^{73}$Ge, $^{74}$Ge, and $^{76}$Ge, with corresponding natural abundances of 20.38, 27.31, 7.76, 36.72, and 7.83\%.~\cite{de2003atomic} The phonon frequency in cm$^{-1}$ for this framework is given by $\omega = \sqrt{K/4\mu\pi^2c^2}$, where $K$ is the force constant, $\mu$ is the reduced mass, and $c$ is the speed of light. Assuming that the force constant $K$ is fixed by the electronic bonding and is the same for all isotopes, the isotope substitution will only modify the reduced mass $\mu$. The statistical relative intensities are computed for each isotopic pair based on natural abundances, which follow a binomial distribution for random isotopic pairs in the lattice. For a homonuclear pair, intensity $I_{ii} = a(i)^2$, and for a heteronuclear pair,  $I_{ij} = 2\times a(i)\times a(j)$. Different combinations of isotopes contribute to the same frequency, for example, a combination of $^{74}$Ge$-^{74}$Ge and $^{72}$Ge$-^{76}$Ge gives the same frequency of $\sim$292.5\,cm$^{-1}$, therefore cumulative intensity is plotted (see Fig.~\ref{fig_CGT_Geisotopes}b). The calculated spectrum using isotopes of Ge for CrGeTe$_3$ reproduces the experimentally reported Raman spectrum in Ref.~\citenum{krasucki2025spin}, confirming their origin from the isotope mass effect (see Fig.~\ref{fig_CGT_Geisotopes}a). We have also calculated the $A_g(5)$ phonon spectrum for different isotopes of Ge using MLMD simulations (see Fig.~\ref{fig_CGT_Geisotopes}c) and the response is similar to that of the simple spring mass model.

We extend the above approach to CrSiTe$_3$, a structurally similar compound to CrGeTe$_3$, which exhibits a frequency comb-like feature up to 100\,K.~\cite{chen2025spontaneously} Naturally occurring Si isotopes are $^{28}$Si, $^{29}$Si, and $^{30}$Si, with natural abundances 92.2, 4.68, and  3.09\%.~\cite{de2003atomic} The combination of $^{28}$Si and $^{29}$Si results in a frequency shift of $\sim$4.3\,cm$^{-1}$ relative to two $^{28}$Si atoms, while the combination of $^{28}$Si and $^{30}$Si shifts by $\sim$4.6 from the combination of $^{28}$Si and $^{29}$Si pair (see Fig.~\ref{fig_CGT_Geisotopes}d). Other combinations, such as those involving both $^{29}$Si and $^{30}$Si isotopes, are not seen due to their relatively low abundances. This also explains why CrSiTe$_3$ exhibits a much larger frequency spacing than CrGeTe$_3$; the lighter Si isotopes produce a larger change in reduced mass, and therefore a larger frequency shift. We note that another recent study by Belojica \emph{et al.}~\cite{belojica2026phonon} has misinterpreted the features in the Raman spectrum of InSiTe$_3$ to be originating from frequency combs as their calculated spacing from isotopic distribution was $\sim$9 and 17\,cm$^{-1}$; however, the authors did not account for the $^{28}$Si/$^{29}$Si and $^{28}$Si/$^{30}$Si isotope combination peaks in their calculations.

\subsubsection{WSe$_2$} \label{WSe2isotopes}

\begin{figure}
    \centering
    \includegraphics[width=0.8\linewidth]{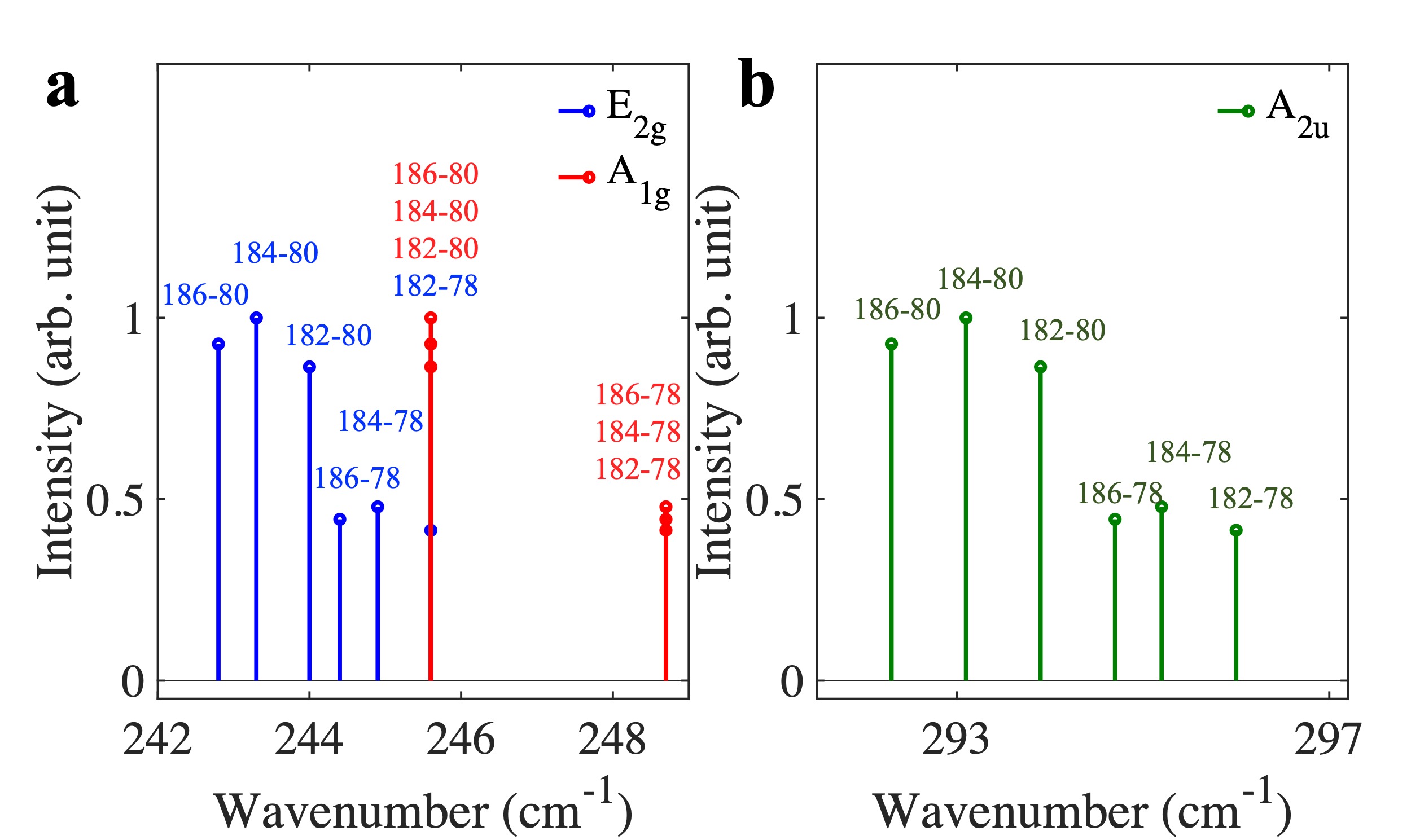}
    \caption[WSe2 isotopes effect on spectrum]{Phonon frequencies corresponding to different combinations of isotopes of W and Se for the $E_{2g}$, $A_{1g}$ and $A_{2u}$ modes. For example, in 186-80, the first number 186 corresponds to the mass of the W isotope, and 80 corresponds to the mass of the Se isotope.}
    \label{fig_WSe2isotopes}
\end{figure}

Selenium have two primary isotopes $^{80}$Se and $^{78}$Se with natural abundances of 49.61 and 23.77\%, while tungsten has three primary isotopes - $^{186}$W, $^{184}$W, and $^{182}$W with natural abundances of 28.43, 30.64, and 26.50\%. The phonon frequencies are calculated for different combinations of these masses, with intensities proportional to their natural abundances (see Fig.~\ref{fig_WSe2isotopes}). The labeling of frequencies for different isotope combinations is also given. The spread in phonon frequencies of the $E_{2g}$ mode, if the sample was synthesized from natural abundances of W and Se, will be from 242.8 to 245.6\,cm$^{-1}$, which is smaller than the spread in the peaks observed in the main text Fig.~\ref{fig_WSe2_potential}(c) and Fig.~\ref{fig_E2g_A2u_potential}(c). Moreover, the intensities of the peaks arising from the natural abundances of W and Se isotopes are comparable, in contrast to the first adjacent peak in the phonon frequency combs, whose intensity is only 16\% of the main peak, thereby ruling out the origin of the combs from the presence of isotopes. Similarly, the $A_{1g}$ will have two frequencies separated by $\sim$3.1\,cm$^{-1}$ with the sideband appearing on the higher-energy side and not interfering with the combs of the $E_{2g}$ mode.

\subsection{Exclusion of thermal hot band mechanism} \label{sectionhotband}

To assess whether the satellite feature adjacent to the $E_{2g}$ fundamental peak could arise from a thermally populated vibrational level (hot band), we consider the standard Boltzmann population for an anharmonic oscillator. The relative intensity of the $n$-th hot band, originating from the population of the $n$ level, scales as

\begin{equation} \label{eq:hotband}
\frac{I_n}{I_0} = (n+1)\exp\left(-\frac{n\hbar\omega}{k_BT}\right) \times \exp\left(-\frac{(n-1)\hbar \Delta \omega}{k_BT}\right),
\end{equation}
where $\hbar\omega \sim 244~\mathrm{cm^{-1}}$ is the fundamental mode energy.  Evaluating Eq.~\ref{eq:hotband} at 50\,K for the first hot band ($n=1$) leads to

\begin{equation} \label{eq:hotband_n1}
\frac{I_1}{I_0} =2 e^{\left(-\frac{244}{34.75}\right) }=2 e^{-7.02} \approx 1.8\times10^{-3},
\end{equation}
which is vanishingly small and three orders of magnitude smaller than the fundamental peak, whereas our SED data show an adjacent peak with an intensity ratio of 16\%, thereby ruling out the possibility of hot bands.


\clearpage

\subsection{Raman tensor} \label{Ramantensorcalculation}
\begin{figure}
    \centering
    \includegraphics[width=1\linewidth]{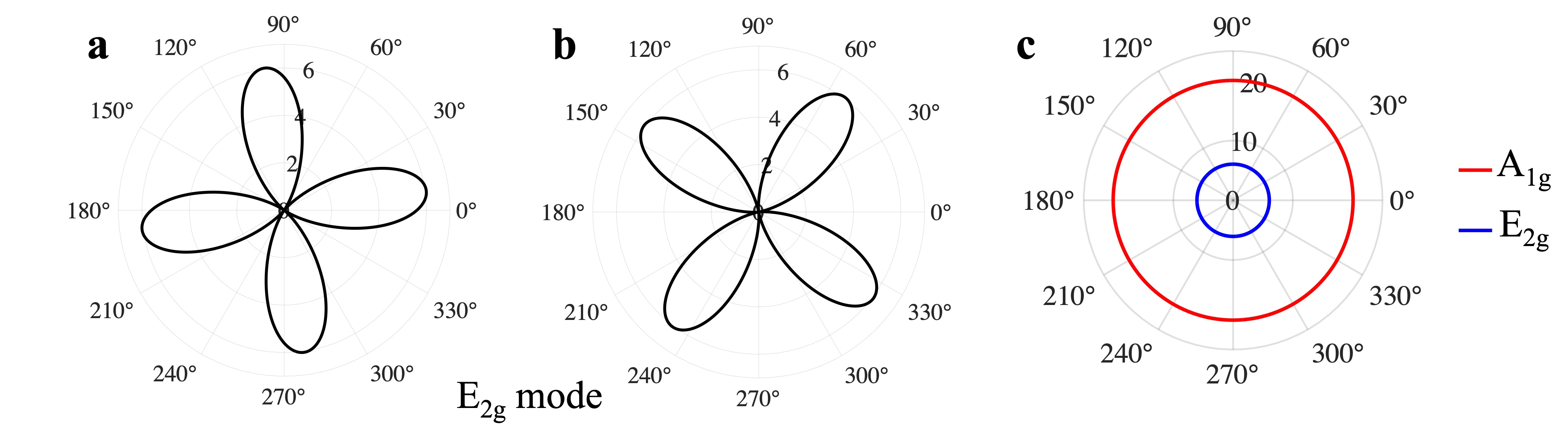}
    \caption[WSe2-Intensity with theta]{(a,b) Variation of Raman intensity with incident angle theta for the doubly degenerate $E_{2g}$ mode of WSe$_2$. The maximum intensity angle of the degenerate $E_{2g}$ mode in the polar plot differs due to differences in the Raman tensor. (c) Comparison of intensities of $A_{1g}$ and $E_{2g}$ modes. Here, the $E_{2g}$ mode intensity is the sum of the two degenerate $E_{2g}$ modes shown in panels (a) and (b).}
    \label{fig_WSe2_Ramanintensitieswiththeta}
\end{figure}

We simulated the Raman intensities of the $E_{2g}$ and $A_{1g}$ modes for various parallel incident polarizations using Raman activity tensors. Raman activity tensors for the $E_{2g}$ and $A_{1g}$ modes of WSe$_2$ are calculated by the change in the polarizability tensor ($\alpha$) along the phonon mode eigenvectors using Phonopy-Spectroscopy.~\cite{Phonopy,skelton2017lattice} We find that the $A_{1g}$ mode has $\sim$3 times the intensity of the $E_{2g}$ mode for all polarization angles after accounting for the degeneracy of the $E_{2g}$ mode (see Fig.~\ref{fig_WSe2_Ramanintensitieswiththeta}). This intensity ratio is consistent with the Raman intensity calculation of WS$_2$.~\cite{ding2020raman}

\clearpage
\newpage

\addcontentsline{toc}{chapter}{References} 


\begin{thebibliography}{61}

\bibitem{pupeza2021extreme}
Ioachim Pupeza, Chuankun Zhang, Maximilian H{\"o}gner, and Jun Ye.
\newblock Extreme-ultraviolet frequency combs for precision metrology and
  attosecond science.
\newblock {\em Nature Photonics}, 15(3):175--186, 2021.

\bibitem{kippenberg2011microresonator}
Tobias~J Kippenberg, Ronald Holzwarth, and Scott~A Diddams.
\newblock Microresonator-based optical frequency combs.
\newblock {\em Science}, 332(6029):555--559, 2011.

\bibitem{udem2002optical}
Th~Udem, Ronald Holzwarth, and Theodor~W H{\"a}nsch.
\newblock Optical frequency metrology.
\newblock {\em Nature}, 416(6877):233--237, 2002.

\bibitem{cundiff2003colloquium}
Steven~T Cundiff and Jun Ye.
\newblock Colloquium: Femtosecond optical frequency combs.
\newblock {\em Reviews of Modern Physics}, 75(1):325, 2003.

\bibitem{gohle2005frequency}
Christoph Gohle, Thomas Udem, Maximilian Herrmann, Jens Rauschenberger, Ronald
  Holzwarth, Hans~A Schuessler, Ferenc Krausz, and Theodor~W H{\"a}nsch.
\newblock A frequency comb in the extreme ultraviolet.
\newblock {\em Nature}, 436(7048):234--237, 2005.

\bibitem{jones2005phase}
R~Jason Jones, Kevin~D Moll, Michael~J Thorpe, and Jun Ye.
\newblock Phase-coherent frequency combs in the vacuum ultraviolet via
  high-harmonic generation inside a femtosecond enhancement cavity.
\newblock {\em Physical Review Letters}, 94(19):193201, 2005.

\bibitem{keilmann2004time}
Fritz Keilmann, Christoph Gohle, and Ronald Holzwarth.
\newblock Time-domain mid-infrared frequency-comb spectrometer.
\newblock {\em Optics Letters}, 29(13):1542--1544, 2004.

\bibitem{huang2008spectral}
Chen-Bin Huang, Zhi Jiang, DanielE Leaird, Jose Caraquitena, and AndrewM
  Weiner.
\newblock Spectral line-by-line shaping for optical and microwave arbitrary
  waveform generations.
\newblock {\em Laser \& Photonics Reviews}, 2(4):227--248, 2008.

\bibitem{cao2014phononic}
LS~Cao, DX~Qi, RW~Peng, Mu~Wang, and P~Schmelcher.
\newblock Phononic frequency combs through nonlinear resonances.
\newblock {\em Physical Review Letters}, 112(7):075505, 2014.

\bibitem{ganesan2017phononic}
Adarsh Ganesan, Cuong Do, and Ashwin Seshia.
\newblock Phononic frequency comb via intrinsic three-wave mixing.
\newblock {\em Physical Review Letters}, 118(3):033903, 2017.

\bibitem{ganesan2018phononic}
Adarsh Ganesan, Cuong Do, and Ashwin Seshia.
\newblock Phononic frequency comb via three-mode parametric resonance.
\newblock {\em Applied Physics Letters}, 112(2), 2018.

\bibitem{ganesan2018phononicprb}
Adarsh Ganesan, Cuong Do, and Ashwin Seshia.
\newblock Excitation of coupled phononic frequency combs via two-mode
  parametric three-wave mixing.
\newblock {\em Physical Review B}, 97:014302, 2018.

\bibitem{xu2025phononic}
Nan Xu, Zi-Jian Zhang, Sheng-Jie Xue, Tong Li, Qiang Zhou, You Wang, Hai-Zhi
  Song, Ke~Zhang, Konstantin Arutyunov, Xin-He Wang, et~al.
\newblock Phononic frequency comb in carbon nanotube mechanical resonators at
  very high frequency band.
\newblock {\em Frontiers of Physics}, 20(3):032202, 2025.

\bibitem{wang2007thermal}
Lei Wang and Baowen Li.
\newblock Thermal logic gates: computation with phonons.
\newblock {\em Physical Review Letters}, 99(17):177208, 2007.

\bibitem{middlemiss2016measurement}
RP~Middlemiss, Antonio Samarelli, DJ~Paul, James Hough, Sheila Rowan, and
  GD~Hammond.
\newblock Measurement of the earth tides with a mems gravimeter.
\newblock {\em Nature}, 531(7596):614--617, 2016.

\bibitem{vahala2009phonon}
Kerry Vahala, Maximilian Herrmann, Sebastian Kn{\"u}nz, Valentin Batteiger,
  Guido Saathoff, TW~H{\"a}nsch, and Th~Udem.
\newblock A phonon laser.
\newblock {\em Nature Physics}, 5(9):682--686, 2009.

\bibitem{rangwala2026spontaneous}
Murtaza Rangwala and Adarsh Ganesan.
\newblock Spontaneous symmetry breaking and collective {Higgs-Goldstone}
  dynamics in solid-state phononic frequency combs.
\newblock {\em arXiv preprint arXiv:2602.07462}, 2026.

\bibitem{chen2025spontaneously}
Lebing Chen, Gaihua Ye, Cynthia Nnokwe, Xing-Chen Pan, Katsumi Tanigaki,
  Guanghui Cheng, Yong~P Chen, Jiaqiang Yan, David~G Mandrus, Andres~E
  Llacsahuanga~Allcca, et~al.
\newblock Spontaneously formed phonon frequency combs in van der {Waals} solid
  {CrGeTe$_3$} and {CrSiTe$_3$}.
\newblock {\em Nature Communications}, 16(1):5795, 2025.

\bibitem{krasucki2025spin}
Grzegorz Krasucki, Katarzyna Olkowska-Pucko, Tomasz Wozniak, Mihai~I Sturza,
  Holger Kohlmann, Maciej~R Molas, et~al.
\newblock Spin-phonon coupling and isotope-related pseudo-molecule vibrations
  in layered {Cr$_2$Ge$_2$Te$_6$} ferromagnet.
\newblock {\em arXiv preprint arXiv:2510.01881}, 2025.

\bibitem{mayur1994fine}
AJ~Mayur, M~Dean Sciacca, MK~Udo, AK~Ramdas, K~Itoh, J~Wolk, and EE~Haller.
\newblock Fine structure of the asymmetric stretching vibration of dispersed
  oxygen in monoisotopic germanium.
\newblock {\em Physical Review B}, 49(23):16293, 1994.

\bibitem{belojica2026phonon}
Tea Belojica, Jovan Blagojevi{\'c}, Sanja Djurdji{\'c}~Mijin, Andrijana
  {\v{S}}olaji{\'c}, Jelena Pe{\v{s}}i{\'c}, Emil~S Bozin, Bojana
  Vi{\v{s}}i{\'c}, Yu~Liu, Cedomir Petrovic, Zoran~V Popovi{\'c}, et~al.
\newblock Phonon frequency comb close to an isolated {Einstein} mode in
  {InSiTe$_3$}.
\newblock {\em Scientific reports}, 2026.

\bibitem{Born1988}
M.~Born and K.~Huang.
\newblock {\em Dynamical theory of crystal lattices}.
\newblock Clarendon Press, Oxford, 1988.

\bibitem{Bruesch}
P.~Br\"uesch.
\newblock {\em Phonons: Theory and Experiments I -- Lattice Dynamics and Models
  of Interatomic Forces}.
\newblock Vol. 34, Springer Series in Solid-State Sciences, New York, 1982.

\bibitem{Dove1993}
M.~T. Dove.
\newblock {\em Introduction to Lattice Dynamics}.
\newblock Cambridge University Press, Cambridge, 1993.

\bibitem{wu2008ultrafast}
Alexander~Q Wu, Xianfan Xu, and Rama Venkatasubramanian.
\newblock Ultrafast dynamics of photoexcited coherent phonon in {Bi$_2$Te$_3$}
  thin films.
\newblock {\em Applied Physics Letters}, 92(1), 2008.

\bibitem{liu2025phononic}
Rumeng Liu and Guangfei Zhu.
\newblock Phononic frequency combs in twisted bilayer van der {Waals} material
  resonators.
\newblock {\em Journal of Applied Physics}, 138(14), 2025.

\bibitem{carteaux1995crystallographic}
V~Carteaux, Dominique Brunet, Guy Ouvrard, and Gilles Andr{\'e}.
\newblock Crystallographic, magnetic and electronic structures of a new layered
  ferromagnetic compound {Cr$_2$Ge$_2$Te$_6$}.
\newblock {\em Journal of Physics: Condensed Matter}, 7(1):69, 1995.

\bibitem{zhang2019first}
B.~H. Zhang, Y.~S. Hou, Z~Wang, and R.~Q. Wu.
\newblock First-principles studies of spin-phonon coupling in monolayer
  {Cr$_2$Ge$_2$Te$_6$}.
\newblock {\em Physical Review B}, 100(22):224427, 2019.

\bibitem{bansal2020magnetically}
Dipanshu Bansal, Jennifer~L. Niedziela, Stuart Calder, Tyson Lanigan-Atkins,
  Ryan Rawl, Ayman~H. Said, Douglas~L. Abernathy, Alexander~I. Kolesnikov,
  Haidong Zhou, and Olivier Delaire.
\newblock Magnetically driven phonon instability enables the metal--insulator
  transition in {h-FeS}.
\newblock {\em Nature Physics}, 16:669--675, 2020.

\bibitem{tadano2014anharmonic}
Terumasa Tadano, Yoshihiro Gohda, and Shinji Tsuneyuki.
\newblock Anharmonic force constants extracted from first-principles molecular
  dynamics: applications to heat transfer simulations.
\newblock {\em Journal of Physics: Condensed Matter}, 26(22):225402, 2014.

\bibitem{korotin2015calculation}
Dm.~M. Korotin, V.~V. Mazurenko, V.~I. Anisimov, and S.~V. Streltsov.
\newblock Calculation of exchange constants of the {Heisenberg} model in
  plane-wave-based methods using the {Green's} function approach.
\newblock {\em Physical Review B}, 91:224405, 2015.

\bibitem{jayakrishnan2025coherent}
SS~Jayakrishnan and Dipanshu Bansal.
\newblock Coherent phonon excitation induced evolution of spin dynamics and
  spin-phonon coupling in yttrium orthochromite.
\newblock {\em Physical Review B}, 112(21):214419, 2025.

\bibitem{ss2025effect}
Jayakrishnan SS and Dipanshu Bansal.
\newblock Effect of spin-phonon coupling on phonons and magnons in the
  antiferromagnet {NiO}.
\newblock {\em Physical Review B}, 111(10):104306, 2025.

\bibitem{tian2016magneto}
Yao Tian, Mason~J Gray, Huiwen Ji, Robert~J Cava, and Kenneth~S Burch.
\newblock Magneto-elastic coupling in a potential ferromagnetic {2D} atomic
  crystal.
\newblock {\em 2D Materials}, 3(2):025035, 2016.

\bibitem{pataniya2020low}
Pratik~M Pataniya, Mohit Tannarana, Chetan~K Zankat, Sanjay~A Bhakhar, Som
  Narayan, Gunvant~K Solanki, Kirit~D Patel, Prafulla~K Jha, and Vivek~M
  Pathak.
\newblock Low-temperature {Raman} investigations and photoresponse of a
  detector based on high-quality {WSe$_2$} crystals.
\newblock {\em The Journal of Physical Chemistry C}, 124(4):2251--2257, 2020.

\bibitem{chen2015helicity}
Shao-Yu Chen, Changxi Zheng, Michael~S Fuhrer, and Jun Yan.
\newblock Helicity-resolved {Raman} scattering of {MoS$_2$, MoSe$_2$, WS$_2$,
  and WSe$_2$} atomic layers.
\newblock {\em Nano Letters}, 15(4):2526--2532, 2015.

\bibitem{kresse1996efficient}
Georg Kresse and J{\"u}rgen Furthm{\"u}ller.
\newblock Efficient iterative schemes for \emph{ab initio} total-energy
  calculations using a plane-wave basis set.
\newblock {\em Physical Review B}, 54(16):11169, 1996.

\bibitem{kresse1996efficiency}
Georg Kresse and J{\"u}rgen Furthm{\"u}ller.
\newblock Efficiency of \emph{ab initio} total energy calculations for metals
  and semiconductors using a plane-wave basis set.
\newblock {\em Computational Materials Science}, 6(1):15--50, 1996.

\bibitem{kresse1993ab}
Georg Kresse and J{\"u}rgen Hafner.
\newblock \emph{Ab initio} molecular dynamics for liquid metals.
\newblock {\em Physical Review B}, 47(1):558, 1993.

\bibitem{wang2023magnetic}
Xiaotian Wang, Xiao-Ping Li, Jianghua Li, Chengwu Xie, Jianhua Wang, Hongkuan
  Yuan, Wenhong Wang, Zhenxiang Cheng, Zhi-Ming Yu, and Gang Zhang.
\newblock Magnetic second-order topological insulator: an experimentally
  feasible {2D} {CrSiTe$_3$}.
\newblock {\em Advanced Functional Materials}, 33(49):2304499, 2023.

\bibitem{wang2018deepmd}
Han Wang, Linfeng Zhang, Jiequn Han, et~al.
\newblock Deepmd-kit: A deep learning package for many-body potential energy
  representation and molecular dynamics.
\newblock {\em Computer Physics Communications}, 228:178--184, 2018.

\bibitem{plimpton1995fast}
Steve Plimpton.
\newblock Fast parallel algorithms for short-range molecular dynamics.
\newblock {\em Journal of Computational Physics}, 117(1):1--19, 1995.

\bibitem{thomas2010predicting}
John~A Thomas, Joseph~E Turney, Ryan~M Iutzi, Cristina~H Amon, and Alan~JH
  McGaughey.
\newblock Predicting phonon dispersion relations and lifetimes from the
  spectral energy density.
\newblock {\em Physical Review B—Condensed Matter and Materials Physics},
  81(8):081411, 2010.

\bibitem{gupta2025molecular}
Mayanak~K Gupta.
\newblock A molecular dynamics postprocessing tool for analyzing the structure
  and dynamics of materials.
\newblock {\em Computer Physics Communications}, page 109982, 2025.

\bibitem{gupta2023distinct}
Mayanak~K Gupta, Sajan Kumar, Ranjan Mittal, Sanjay~K Mishra, Stephane Rols,
  Olivier Delaire, Arumugum Thamizhavel, PU~Sastry, and Samrath~L Chaplot.
\newblock Distinct anharmonic characteristics of phonon-driven lattice thermal
  conductivity and thermal expansion in bulk {MoSe$_2$} and {WSe$_2$}.
\newblock {\em Journal of Materials Chemistry A}, 11(40):21864--21873, 2023.

\bibitem{cai2022monolayer}
Qingan Cai, Bin Wei, Qiyang Sun, Ayman~H Said, and Chen Li.
\newblock Monolayer-like lattice dynamics in bulk {WSe$_2$}.
\newblock {\em Materials Today Physics}, 28:100856, 2022.

\bibitem{QE-2017}
P.~Giannozzi, O.~Andreussi, T.~Brumme, O.~Bunau, M.~Buongiorno Nardelli,
  M.~Calandra, R.~Car, C.~Cavazzoni, D.~Ceresoli, M.~Cococcioni, N.~Colonna,
  I.~Carnimeo, A.~Dal Corso, S.~de~Gironcoli, P.~Delugas, R.~A.~DiStasio Jr,
  A.~Ferretti, A.~Floris, G.~Fratesi, G.~Fugallo, R.~Gebauer, U.~Gerstmann,
  F.~Giustino, T.~Gorni, J.~Jia, M.~Kawamura, H-Y. Ko, A.~Kokalj,
  E.~Küçükbenli, M.~Lazzeri, M.~Marsili, N.~Marzari, F.~Mauri, N.~L. Nguyen,
  H-V. Nguyen, A.~Otero de-la Roza, L.~Paulatto, S.~Poncé, D.~Rocca,
  R.~Sabatini, B.~Santra, M.~Schlipf, A.~P. Seitsonen, A.~Smogunov, I.~Timrov,
  T.~Thonhauser, P.~Umari, N.~Vast, X.~Wu, and S.~Baroni.
\newblock Advanced capabilities for materials modelling with {QUANTUM
  ESPRESSO}.
\newblock {\em Journal of Physics: Condensed Matter}, 29(46):465901, 2017.

\bibitem{QE-2009}
Paolo Giannozzi, Stefano Baroni, Nicola Bonini, Matteo Calandra, Roberto Car,
  Carlo Cavazzoni, Davide Ceresoli, Guido~L. Chiarotti, Matteo Cococcioni,
  Ismaila Dabo, Andrea {Dal Corso}, Stefano de~Gironcoli, Stefano Fabris, Guido
  Fratesi, Ralph Gebauer, Uwe Gerstmann, Christos Gougoussis, Anton Kokalj,
  Michele Lazzeri, Layla Martin-Samos, Nicola Marzari, Francesco Mauri,
  Riccardo Mazzarello, Stefano Paolini, Alfredo Pasquarello, Lorenzo Paulatto,
  Carlo Sbraccia, Sandro Scandolo, Gabriele Sclauzero, Ari~P. Seitsonen,
  Alexander Smogunov, Paolo Umari, and Renata~M Wentzcovitch.
\newblock {QUANTUM ESPRESSO}: a modular and open-source software project for
  quantum simulations of materials.
\newblock {\em Journal of Physics: Condensed Matter}, 21(39):395502 (19pp),
  2009.

\bibitem{refPBE}
J.~P. Perdew, K.~Burke, and M.~Ernzerhof.
\newblock Generalized gradient approximation made simple.
\newblock {\em Physical Review Letters}, 77:3865--3868, 1996.

\bibitem{Dudarev_1998}
S.~L. Dudarev, G.~A. Botton, S.~Y. Savrasov, C.~J. Humphreys, and A.~P. Sutton.
\newblock Electron-energy-loss spectra and the structural stability of {N}ickel
  {O}xide: An {LSDA+U} study.
\newblock {\em Physical Review B}, 57:1505, 1998.

\bibitem{fang2018large}
Yimei Fang, Shunqing Wu, Zi-Zhong Zhu, and Guang-Yu Guo.
\newblock Large magneto-optical effects and magnetic anisotropy energy in
  two-dimensional {Cr$_2$Ge$_2$Te$_6$}.
\newblock {\em Physical Review B}, 98(12):125416, 2018.

\bibitem{li2018electronic}
Y.~F. Li, W.~Wang, W.~Guo, C.~Y. Gu, H.~Y. Sun, L.~He, J.~Zhou, Z.~B. Gu, Y.~F.
  Nie, and X.~Q. Pan.
\newblock Electronic structure of ferromagnetic semiconductor
  {Cr$_2$Ge$_2$Te$_6$} by angle-resolved photoemission spectroscopy.
\newblock {\em Physical Review B}, 98:125127, Sep 2018.

\bibitem{pizzi2020wannier90}
Giovanni Pizzi, Valerio Vitale, Ryotaro Arita, Stefan Bl{\"u}gel, Frank
  Freimuth, Guillaume G{\'e}ranton, Marco Gibertini, Dominik Gresch, Charles
  Johnson, Takashi Koretsune, et~al.
\newblock Wannier90 as a community code: new features and applications.
\newblock {\em Journal of Physics: Condensed Matter}, 32(16):165902, 2020.

\bibitem{chen2022anisotropic}
Lebing Chen, Chengjie Mao, Jae-Ho Chung, Matthew~B Stone, Alexander~I
  Kolesnikov, Xiaoping Wang, Naoki Murai, Bin Gao, Olivier Delaire, and
  Pengcheng Dai.
\newblock Anisotropic magnon damping by zero-temperature quantum fluctuations
  in ferromagnetic {CrGeTe$_3$}.
\newblock {\em Nature Communications}, 13(1):4037, 2022.

\bibitem{zhu2021topological}
Fengfeng Zhu, Lichuan Zhang, Xiao Wang, Flaviano~Jos{\'e} Dos~Santos, Junda
  Song, Thomas Mueller, Karin Schmalzl, Wolfgang~F Schmidt, Alexandre Ivanov,
  Jitae~T Park, et~al.
\newblock Topological magnon insulators in two-dimensional van der waals
  ferromagnets {CrSiTe$_3$} and {CrGeTe$_3$}: Toward intrinsic gap-tunability.
\newblock {\em Science Advances}, 7(37):7532, 2021.

\bibitem{zhao2013interlayer}
Yanyuan Zhao, Xin Luo, Hai Li, Jun Zhang, Paulo~T Araujo, Chee~Kwan Gan,
  Jumiati Wu, Hua Zhang, Su~Ying Quek, Mildred~S Dresselhaus, et~al.
\newblock Interlayer breathing and shear modes in few-trilayer {MoS$_2$} and
  {WSe$_2$}.
\newblock {\em Nano letters}, 13(3):1007--1015, 2013.

\bibitem{grzeszczyk2016raman}
Magdalena Grzeszczyk, Katarzyna Go{\l}asa, M~Zinkiewicz, K~Nogajewski,
  Maciej~Roman Molas, Marek Potemski, Andrzej Wysmo{\l}ek, and Adam
  Babi{\'n}ski.
\newblock Raman scattering of few-layers {MoTe$_2$}.
\newblock {\em 2D Materials}, 3(2):025010, 2016.

\bibitem{de2003atomic}
John~R De~Laeter, John~Karl B{\"o}hlke, Paul De~Bievre, H~Hidaka, HS~Peiser,
  KJR Rosman, and PDP Taylor.
\newblock Atomic weights of the elements. review 2000 (iupac technical report).
\newblock {\em Pure and Applied Chemistry}, 75(6):683--800, 2003.

\bibitem{Phonopy}
A~A.~Togo, F.~Oba, and I.~Tanaka.
\newblock First-principles calculations of the ferroelastic transition between
  rutile-type and {CaCl$_2$}-type {SiO$_2$} at high pressures.
\newblock {\em Physical Review B}, 78:134106, 2008.

\bibitem{skelton2017lattice}
Jonathan~M Skelton, Lee~A Burton, Adam~J Jackson, Fumiyasu Oba, Stephen~C
  Parker, and Aron Walsh.
\newblock Lattice dynamics of the tin sulphides {SnS$_2$}, {SnS} and
  {Sn$_2$S$_3$}: vibrational spectra and thermal transport.
\newblock {\em Physical Chemistry Chemical Physics}, 19(19):12452--12465, 2017.

\bibitem{ding2020raman}
Ying Ding, Wei Zheng, Zeguo Lin, Ruinan Zhu, Mingge Jin, Yanming Zhu, and Feng
  Huang.
\newblock Raman tensor of layered {WS$_2$}.
\newblock {\em Science China Materials}, 63(9):1848--1854, 2020.

\end{thebibliography}
\end{document}